# Runtime Instrumentation for Reactive Components (Extended Version)

**Luca Aceto** ✉ ⓘ
Reykjavik University, Reykjavik, Iceland
Gran Sasso Science Institute, L'Aquila, Italy

**Duncan Paul Attard** ✉ ⓘ
University of Glasgow, Glasgow, UK

**Adrian Francalanza** ✉ ⓘ
University of Malta, Msida, Malta

**Anna Ingólfsdóttir** ✉ ⓘ
Reykjavik University, Reykjavik, Iceland

### Abstract

Reactive software calls for instrumentation methods that uphold the reactive attributes of systems. Runtime verification imposes another demand on the instrumentation, namely that the trace event sequences it reports to monitors are *sound*—that is, they reflect actual executions of the system under scrutiny. This paper presents RIARC, a novel decentralised instrumentation algorithm for outline monitors meeting these two demands. The asynchronous setting of reactive software complicates the instrumentation due to potential trace event loss or reordering. RIARC overcomes these challenges using a next-hop IP routing approach to rearrange and report events soundly to monitors.

RIARC is validated in two ways. We subject its corresponding implementation to rigorous systematic testing to confirm its correctness. In addition, we assess this implementation via extensive empirical experiments, subjecting it to large realistic workloads to ascertain its reactiveness. Our results show that RIARC optimises its memory and scheduler usage to maintain latency feasible for soft real-time applications. We also compare RIARC to inline and centralised monitoring, revealing that it induces comparable latency to inline monitoring in moderate concurrency settings, where software performs long-running, computationally-intensive tasks, such as in Big Data stream processing.

**2012 ACM Subject Classification** Software and its engineering → Software verification and validation

**Keywords and phrases** Runtime instrumentation, decentralised monitoring, reactive systems

**Funding** This work is supported by the Reykjavik University Research Fund, the Doctoral Student Grant (No: 207055) and the MoVeMnt project (No: 217987) under the IRF, and the STARDUST project (No: EP/T014628/1) under the EPSRC.

**Acknowledgements** We thank our anonymous reviewers and the Artifact Evaluation Committee for their feedback. Thanks also to Simon Fowler, Simon J. Gay, and Phil Trinder for their input.

## 1 Introduction

Modern software is generally built in terms of concurrent components that execute without relying on a global clock or shared state [90]. Instead, components interact via non-blocking messaging, creating a loosely-coupled architecture known as a *reactive system* [8, 97], which:

- responds in a timely manner (is *responsive*),
- remains available in the face of failure (is *resilient*),
- reacts to inputs from users or their environment (is *message-driven*), and
- grows and shrinks to accommodate varying computational loads (is *elastic*).

The real-world behaviour of reactive systems is hard to understand statically, and *monitoring* is used to inspect their operation at *runtime*, *e.g.* for debugging [114], security checking [63], profiling [79], resource usage analysis [37], *etc.* This paper considers runtime verification (RV), an application of monitoring used to detect whether the *current* execution of a system under scrutiny (SuS) deviates from its correct behaviour [15, 74, 21]. A RV monitor is a *sequence recogniser* [130, 104]: a state machine that incrementally analyses a *finite* fragment of the runtime information exhibited by a SuS to reach an *irrevocable* verdict (see [6, 5] for details).

*Instrumentation* lies at the core of runtime monitoring [73, 21, 65]. It is the mechanism by which runtime information from a SuS is extracted and reported to monitors as a stream of system events called a *trace*. Software is typically instrumented in one of two ways. Inline instrumentation, or *inlining*, modifies the SuS by injecting tracing instructions at specific joinpoints, *e.g.* using AspectJ [93] or BCEL [54]. Outline instrumentation, or *outlining*, uses an external tracing infrastructure to gather events, *e.g.* LTTng [56] or OpenJ9 [59], thereby treating the SuS as a *black box*. A key requirement setting RV apart from monitoring, *e.g.*, telemetry [88] or profiling [128, 26], is that the instrumentation must report *sound traces*.

▶ **Definition 1** (Sound traces). *A finite trace $T$ is* sound *w.r.t. a system component $P$ iff it is*
1. Complete. *$T$ contains* all *the events exhibited by $P$ so far,* and
2. Consistent. *The event sequence in $T$ reflects the order these occur* locally *at $P$.*    ◀

Traces that violate this soundness invariant are unfit for RV, as omitted, spurious, or out-of-sequence events incorrectly characterise the system behaviour, *nullifying* the verdicts that monitors flag [21, 52]. Reactive software imposes another requirement: that the instrumentation *safeguards* the responsive, resilient, message-driven, and elastic attributes of the SuS. This necessitates an instrumentation method that is itself *reactive*, such that it:
1. does not hamper the SuS by inducing unfeasible runtime overhead (is responsive),
2. permits monitors to fail independently of SuS components (is resilient),
3. reacts to trace events without blocking the SuS (is message-driven), and
4. grows and shrinks in proportion to the size of the SuS (is elastic).

**Limitations of current RV instrumentation methods**    State-of-the-art RV tools use instrumentation methods that do not satisfy *all* of the conditions 1–4 above. This renders them inapplicable to reactive software; see [65, Tables 3 and 4] for details. Many approaches, including [24, 31, 49, 78, 113, 129, 134, 17], assume systems with a *fixed* architecture where the number of components remains constant at runtime, failing to meet condition 4. Works foregoing the assumption of a fixed system size, such as [45, 94, 61, 60, 25, 31, 71, 3], inline the SuS with monitors *statically*. Inlining monitors pre-deployment inherently accommodates systems that grow and shrink (condition 4) as a by-product of the embedded monitor code that executes on the same thread of system components; see fig. 1a. This scheme, however, has shortcomings that make it less suited to reactive software. Recent studies [21, 52] observe that the lock-step execution of the SuS and monitors can impair the operation of the instrumented system, *e.g.* slow runtime analyses manifest as high latencies [38], and faulty monitors may break the system [72], which do not meet conditions 1 and 2 (*e.g.* $M_Q$ in fig. 1a). Other works [46, 14] argue that errors, such as deadlocks or component crashes, are hard to detect since the monitoring logic shares the runtime thread of the affected component. Entwining the execution of the SuS and monitors may also diminish the scalability, performance, and resource usage efficiency of the monitored system because inlined monitor code cannot be run on separate threads [11]. Lastly, inlining is *incompatible* with unmodifiable software, such as closed-source components (*e.g.* $R$ in figs. 1a–1c), making outlining the only alternative.

Outline instrumentation *can* address the limitations of inlining by isolating the SuS and its monitors (works [45, 38, 39] that view externalised monitors as 'outline' embed tracing code to extract events from the SuS, subjecting them to the cons of inlining). The latest survey on decentralised RV [74, Tables 1 and 2] establishes that outlining-based tools, *e.g.* [50, 16, 17, 75, 38, 39, 132, 66], are variations on *centralised* instrumentation. In this set-up, events exhibited by SuS components are funnelled through a *global* trace buffer (*e.g.* $\kappa_{\{P,Q,R\}}$ in fig. 1b) that a singleton monitor can analyse asynchronously, meeting condition 3. Yet, the central buffer introduces contention and sacrifices the scalability of the SuS [10], violating condition 4. Centralised architectures are prone to single point of failures (SPOFs) [97, 96] (violating condition 2), which is not ideal for monitoring medium-scale reactive systems.

**Contribution** We propose RIARC, a *decentralised* instrumentation algorithm for outline monitors that overcomes the above shortcomings, fulfilling conditions 1–4. Outline monitors minimise latency effects due to slow trace event analyses associated with inlining (meeting condition 1). While RIARC does not handle monitor failure explicitly, it intrinsically enjoys a modicum of partial failure by isolating the SuS and its decentralised monitor components (meeting condition 2); *e.g.* monitors $M_{\{P\}}$ and $M_{\{Q,R\}}$ in fig. 1c. RIARC uses a tracing infrastructure to obtain system events passively without modifying the SuS (meeting condition 3). The algorithm equips each isolated monitor with a *local* trace buffer, using it to report events based on the SuS components a monitor is tasked to analyse (*e.g.* buffers $\kappa_{\{P\}}$ and $\kappa_{\{Q,R\}}$ in fig. 1c). RIARC reorganises its instrumentation set-up to reflect dynamic changes in the SuS. It reacts to specific events in traces to instrument monitors for new SuS components and to remove redundant monitors when it detects graceful or abnormal component terminations. This enables RIARC to grow and shrink the verification set-up on demand (meeting condition 4). Given the challenges in fulfilling the conditions 1–4, we scope our work to settings where communication is reliable (*i.e.,* no message corruption, duplication, and loss) [58] and Byzantine failures do not arise [99].

To the best of our knowledge, the approach RIARC advocates is novel. One reason why outlining has never been adopted for decentralising monitors are the onerous conditions 1–4 imposed by reactive software. Utilising non-invasive tracing makes our set-up necessarily *asynchronous*. At the same time, this complicates the instrumentation, which must ensure trace soundness (def. 1), notwithstanding the inherent phenomena arising from the concurrent execution of the SuS and monitors, *e.g.* trace event reordering and process crashes. Consequently, the second reason is that the overhead incurred to uphold this invariant—in addition to scaling the verification set-up as the SuS executes—is perceived as prohibitive when compared to inlining. This opinion is often reinforced when the viability of outline instrumentation is predicated on empirical criteria tied to monolithic, batch-style programs, that *may not* apply to reactive software (*e.g.* percentage slowdown); *e.g.* see [100, 117, 116, 47, 46, 124, 30, 101].

This paper shows how instrumenting outline monitors under conditions 1–4 can be achieved using a decentralised approach that guarantees def. 1, *and* with overheads considered feasible for typical soft real-time reactive systems. Concretely, we:

**(i)** recall the benefits of the actor model of computation [85, 9] for building reactive systems and argue how our model of processes and tracers readily maps to that setting, sec. 2;

**(ii)** give a decentralised instrumentation algorithm for outline monitors, detailing how the reactive characteristics of the SuS can be preserved whilst ensuring def. 1, sec. 3;

**(iii)** show the implementability of our algorithm in an actor language and systematically validate the correctness of its corresponding implementation w.r.t. def. 1 by exhaustively inducing interleaved executions for a selection of instrumented systems, sec. 4;

**(iv)** back up the feasibility of the implemented algorithm via a comprehensive empirical study that uses various workload configurations surpassing the state of the art, showing that the induced overhead minimally impacts the reactive attributes of the SuS, sec. 5.

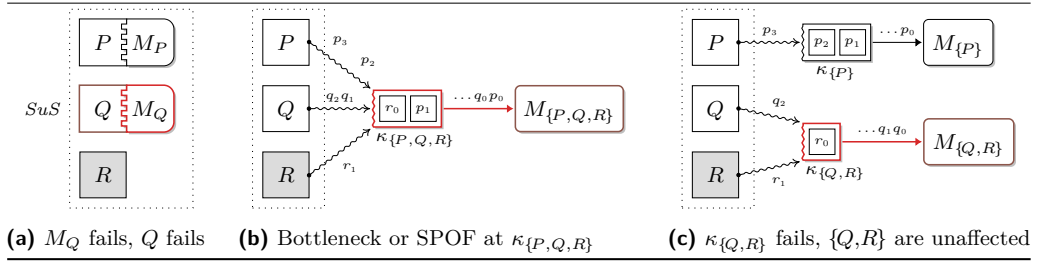

**(a)** $M_Q$ fails, $Q$ fails  **(b)** Bottleneck or SPOF at $\kappa_{\{P,Q,R\}}$  **(c)** $\kappa_{\{Q,R\}}$ fails, $\{Q,R\}$ are unaffected

▪ **Figure 1** $P,Q,R$ instrumented in inline (*left*), centralised (*middle*) and decentralised (*right*) modes

## 2 A computational model for reactive systems

The actor model [85, 9] emerged as *the* paradigm to design and build reactive systems [33]. *Actors*—the units of decomposition in this model—are abstractions of concurrent entities that share no mutable memory with other actors. Instead, actors interact through asynchronous message passing and alter their internal state based on the messages they consume. Asynchronous communication decouples actors spatially and temporally, which fully isolates system components and establishes the foundation for resiliency and elasticity [32, 97]. Each actor is equipped with an incoming message buffer called the *mailbox*, from which messages deposited by other actors can be selectively read. Besides sending and receiving messages, actors can *spawn* other actors. Actors in a system are addressable by their unique process identifier (PID), which they use to engage in directed, *point-to-point* communication. This idea of addressability is central to the actor model: it enables reasoning about decentralised computation, as the identity of components or messages can be propagated through a system and used in handling complex tasks, such as process registration and failure recovery [33]. As is often the case in decentralised computations, we assume that messages exchanged between pairs of processes are always received in the order in which they have been sent [43].

Frameworks, notably Erlang [11], Elixir [91], Akka [1] for Scala [120], along with others [123, 139], instantiate the actor model. We adopt Erlang since its ecosystem is specifically engineered for highly-concurrent, soft real-time reactive systems [140, 12, 44]. The Erlang virtual machine (EVM) implements actors as lightweight processes. It employs *per process* garbage collection that, unlike the JVM, does not subject the virtual machine to global unpredictable pauses [89, 119]. This factor minimises the impact on the soft real-time properties of a system *and* is also crucial to the empirical evaluation of sec. 5 since it stabilises the variance in our experiments. The EVM exposes a flexible *process tracing* API aimed at reactive software [42]. Erlang provides other components, *e.g.* supervision trees, message queues, *etc.*, for building fault-tolerant distributed applications. While we scope our work to fault-free settings (see sec. 1), adopting Erlang gives us the foundation upon which our work can be naturally extended to address these aspects. Henceforth, we follow the established convention in Erlang literature and use the terms *actor*, *process*, and *component* synonymously.

### 2.1 Process tracing and trace partitioning

Processes in a concurrent system form a *tree*, starting at the *root* process that spawns *child* processes, and so forth[1]. Concurrency induces inherent *partitions* to the execution of the

---

[1] For example, using `spawn()` in Erlang [42] and Elixir [91], `ActorContext.spawn()` in Akka [1], `Actor.createActor()` in Thespian [123], `CreateProcess()` in Windows [111], *etc.*

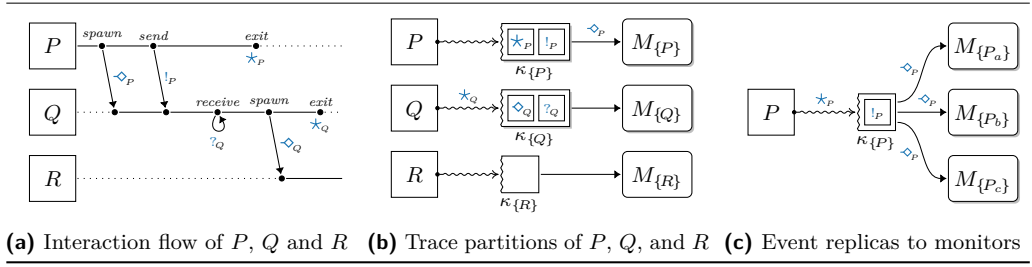

**(a)** Interaction flow of $P$, $Q$ and $R$  **(b)** Trace partitions of $P$, $Q$, and $R$  **(c)** Event replicas to monitors

**Figure 2** SuS with processes $P$, $Q$, and $R$ instrumented with independent monitors

SuS in the form of isolated traces that reflect the *local* behaviour at each process [17]. RIARC exploits this aspect to attain several benefits. First, one can *selectively* specify the SuS processes to be instrumented. The upshot is that fewer trace events need to be gathered, improving *efficiency*. Another benefit of partitioned traces is that each process can be dynamically instrumented, free from assumptions about the number of processes the SuS is expected to have. This makes the RV set-up *elastic*. Lastly, the instrumentation set-up can *partially fail*, as faulty SuS or monitor processes do not imperil the execution of one another.

▶ **Example 2** (Trace partitions). Trace partitions enable RIARC to instrument a system in various arrangements. Fig. 2a depicts an interaction sequence for the execution of the SuS from sec. 1. In this interaction, the root process, $P$, spawns $Q$ and communicates with it, at which point $Q$ spawns process $R$; $P$ and $Q$ eventually terminate. We denote the process *spawning* and *termination* trace events by $\diamond$ and $\star$, and use ! and ? for *send* and *receive* events respectively. The *sound* trace partitions for the processes in fig. 2a are '$\diamond_P.!_P.\star_P$' for $P$, '$?_Q.\diamond_Q.\star_Q$' for $Q$, and the empty trace for $R$.  ◀

A centralised set-up such as that of fig. 1b can be obtained by instrumenting $\{P,Q,R\}$ with one monitor, $M_{\{P,Q,R\}}$, whereas instrumenting the components $\{P\}$ and $\{Q,R\}$ with monitors $M_{\{P\}}$ and $M_{\{Q,R\}}$ gives the decentralised arrangement of fig. 1c. Each of these instrumentation arrangements generates different executions.

▶ **Example 3** (Sound traces). For the case of fig. 1b, RIARC can report to $M_{\{P,Q,R\}}$ *one* of four possible traces '$\diamond_P.!_P.\star_P.?_Q.\diamond_Q.\star_Q$', '$\diamond_P.!_P.?_Q.\star_P.\diamond_Q.\star_Q$', '$\diamond_P.!_P.?_Q.\diamond_Q.\star_P.\star_Q$', or '$\diamond_P.!_P.?_Q.\diamond_Q.\star_Q.\star_P$'. These *sound* traces result from the interleaved execution of processes $P$, $Q$. Any other trace, *e.g.* '$\diamond_P.\star_P.?_Q.\diamond_Q.\star_Q$' or '$\diamond_P.!_P.\star_P.?_Q.\star_Q.\diamond_Q$', is *unsound* since it contradicts the local behaviour at processes $P$ and $Q$ of fig. 2a. The former trace omits the request $!_P$ that $P$ makes to $Q$ (it is *incomplete* w.r.t. $P$), and the latter trace inverts $\diamond_Q$ and $\star_Q$, suggesting that $Q$ spawns $R$ after $Q$ terminates (it is *inconsistent* w.r.t. $Q$).  ◀

▶ **Example 4** (Separate instrumentation). Fig. 2b shows another decentralised set-up, where $P$, $Q$, and $R$ are instrumented separately. In this case, the instrumentation should report to $M_{\{P\}}$, $M_{\{Q\}}$ and $M_{\{R\}}$ the events observed *locally* at each process, as stated in ex. 2.  ◀

RIARC makes two assumptions about process tracing in order to support the instrumentation arrangements shown in figs. 1b, 1c, and 2b:

**A$_1$** *Tracing processes sets.* Tracing can gather events for *sets* of SuS processes, *e.g.* $\kappa_{\{P,Q,R\}}$ in fig. 1b gathers the events of $\{P,Q,R\}$, and $\kappa_{\{Q,R\}}$ in fig. 1c gathers the events of $\{Q,R\}$.
**A$_2$** *Tracing inheritance.* Tracing gathers the events of a SuS process *and* the children it spawns by default to eliminate the risk that trace events from child processes are missed.

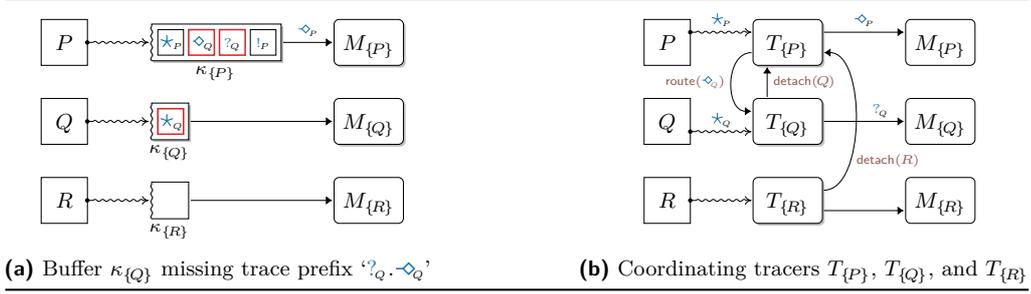

**(a)** Buffer $\kappa_{\{Q\}}$ missing trace prefix '$?_Q.\diamond_Q$'  **(b)** Coordinating tracers $T_{\{P\}}$, $T_{\{Q\}}$, and $T_{\{R\}}$

**Figure 3** Choreographed tracers coordinating to ensure sound traces

We opt for tracing inheritance since it follows established centralised RV monitoring tools, including [16, 41, 50, 113]. In fact, tracing assumptions $A_1$ and $A_2$ mean that centralised set-ups like that of fig. 1b can be obtained just by tracing the root process $P$. Tracing inheritance requires the instrumentation to *intervene* if it needs to channel the events of a child process into a *new* trace partition that is *independent* from that of its parent, *e.g.* as in fig. 1c. In such cases, the instrumentation must first stop tracing the child process, allocate a fresh trace buffer, and resume tracing the child process. The out-of-sync execution of the SuS and instrumentation complicates the creation of these new trace partitions because it can lead to reordered or missed events. This, in turn, would violate trace soundness, def. 1.

We supplement $A_1$ and $A_2$ with the following to keep our exposition in sec. 3 manageable:

**A₃** *Single-process tracing.* Any SuS process can be traced *at most* once at any point in time.
**A₄** *Causally-ordered spawn events.* Tracing gathers the spawn trace event of a parent process before *all* the events of the child process spawned by that parent, *e.g.* if $P$ spawns $Q$, and $Q$ receives, as in fig. 2a, the reported sequence is '$\diamond_P.?_Q$' rather than '$?_Q.\diamond_P$'.

The constraint of tracing assumption $A_3$ is easily overcome by replicating trace events for a process and reporting them to different monitors (*e.g.* the events in the trace partition of process $P$ are replicated to monitors $M_{\{P_a\}}$, $M_{\{P_b\}}$, $M_{\{P_c\}}$ in fig. 2c). Tracing assumption $A_4$ requires trace buffers to reorder $\diamond$ events using the spawner and spawned process information carried by each event before reporting them to monitors. Sec. 3.3 gives more details.

▶ **Example 5** (Unsound traces). Fig. 3a shows one possible configuration that can be reached by our three-process system introduced in fig. 2a, where the trace buffer $\kappa_{\{P\}}$ contains the events for both $P$ and $Q$. The trace in buffer $\kappa_{\{Q\}}$ is unsound, as it inaccurately characterises the local behaviour of process $Q$ (the sound trace for $Q$ should be '$?_Q.\diamond_Q.\star_Q$', not '$\star_Q$'). ◀

RIARC programs trace buffers to coordinate with one another to ensure that sound traces are invariably reported to monitors. We refer to a trace buffer and the coordination logic it encapsulates as a *tracer*. RIARC employs an approach based on *next-hop routing* in IP networks [83, 107] to counteract the effects of trace event reordering and loss by rearranging and forwarding events to different tracers. Fig. 3b conveys our organisation of tracers (refer to fig. 10 in app. A for legend). Sec. 3 details how RIARC dynamically reorganises the tracer choreography and performs next-hop routing.

## 2.2 Modelling decentralised instrumentation

Since RV monitors are passive verdict-flagging machines (refer to sec. 1), they are orthogonal to our instrumentation. We, thus, focus our narrative on tracers and omit monitors, except when relevant in the surrounding context. The model assumes a set of SuS process, $P,Q,R \in \text{PRC}$, and tracer names, $T \in \text{T}_{\text{RC}}$, together with a countable set of PID values to reference processes. We distinguish between SuS and tracer PIDs, which we denote respectively by the sets, $p_{\text{S}}, q_{\text{S}} \in \text{P}_{\text{ID}_{\text{S}}}$ and $p_{\text{T}}, q_{\text{T}} \in \text{P}_{\text{ID}_{\text{T}}}$. The variables $\imath_{\text{S}}$ and $\jmath_{\text{S}}$, and $\imath_{\text{T}}$ and $\jmath_{\text{T}}$ range over PIDs from the corresponding sets $\text{P}_{\text{ID}_{\text{S}}}$ and $\text{P}_{\text{ID}_{\text{T}}}$. We also assume the function signature sets, $f_{\text{S}} \in \text{S}_{\text{IG}_{\text{S}}}$, $f_{\text{T}} \in \text{S}_{\text{IG}_{\text{T}}}$, and, $f_{\text{M}} \in \text{S}_{\text{IG}_{\text{M}}}$, to denote SuS, tracer, and RV monitor functions, together with the variables $\varsigma_{\text{S}}, \varsigma_{\text{T}}$, and $\varsigma_{\text{M}}$ that range over each signature set. New SuS processes are created via the function $\text{spwn}(\varsigma_{\text{S}})$ that accepts the function signature $\varsigma_{\text{S}}$ to be spawned, and returns a fresh PID, $\imath_{\text{S}}$. We overload $\text{spwn}$ to spawn tracer signatures $\varsigma_{\text{T}}$ equivalently, returning corresponding PIDs, $\imath_{\text{T}}$. The function $\text{self}$ obtains the PID of the process invoking it. We write $P$ as shorthand for a singleton process set $\{P\}$ to simplify notation.

RIARC uses three message types, $\tau \in \{\text{evt}, \text{dtc}, \text{rtd}\}$. These determine when to *create* or *terminate* tracer processes, and what trace events to *route* between tracers:

- evt are *trace events* gathered via process tracing,
- dtc are *detach* requests that tracers exchange to reorganise the tracer choreography, and
- rtd are *routing* packets that transport evt or dtc messages forwarded between tracers.

We encode messages $m$ as tuples. Trace event messages, $\langle \text{evt}, \ell, \imath_{\text{S}}, \jmath_{\text{S}}, \varsigma_{\text{S}} \rangle$, comprise the event label $\ell$ that ranges over the SuS events $\diamond$ *(spawn)*, $\star$ *(exit)*, ! *(send)*, and ? *(receive)*. The PID value $\imath_{\text{S}}$ identifies the SuS process exhibiting the trace event, and is defined for *all* events. The SuS PID $\jmath_{\text{S}}$ and function signature $\varsigma_{\text{S}}$ depend on the type of the event. Tbl. 1a catalogues the values defined for each event. We write trace events in their shorthand form, omitting undefined values (denoted by $\bot$), *e.g.* $\langle \text{evt}, \star, \imath_{\text{S}} \rangle$ instead of $\langle \text{evt}, \star, \imath_{\text{S}}, \bot, \bot \rangle$.

Detach request messages have the form $\langle \text{dtc}, \imath_{\text{T}}, \imath_{\text{S}} \rangle$. A tracer with the PID $\imath_{\text{T}}$ uses dtc to request that another tracer *stop* tracing the SuS PID $\imath_{\text{S}}$. Routing packet messages, $\langle \text{rtd}, \imath_{\text{T}}, m \rangle$, move evt and dtc messages between tracers. The PID $\imath_{\text{T}}$ identifies the tracer that embeds the message $m$ into the routing packet and dispatches it to other tracers. Tbl. 1b summarises detach request and routing packet messages.

We reserve the variables $e$, $d$, and $r$ for the messages types evt, dtc, and rtd respectively. Our model uses the suggestive dot notation (.) to index message fields, *e.g.* $m.\tau$ reads the message type, $e.\ell$ reads the trace event label, *etc.* (see tbl. 1). For simplicity, we occasionally write the label $\ell$ in lieu of the full trace event form, *e.g.* we write $\star$ instead of $\langle \text{evt}, \star, \imath_{\text{S}} \rangle$.

| Label $\ell$ | Index | Description ($\imath_{\text{S}}$ and $\jmath_{\text{S}}$ are SuS PIDs) |
|---|---|---|
| $\diamond$ | $e.\imath_{\text{S}}$ | Parent PID spawning new child PID $\jmath_{\text{S}}$ |
| | $e.\jmath_{\text{S}}$ | Child PID spawned by parent PID $\imath_{\text{S}}$ |
| | $e.\varsigma_{\text{S}}$ | Signature $\varsigma_{\text{S}}$ spawned by parent PID $\imath_{\text{S}}$ |
| $\star$ | $e.\imath_{\text{S}}$ | Terminated PID |
| | $e.\jmath_{\text{S}}, e.\varsigma_{\text{S}}$ | *Undefined for exit events* |
| ! | $e.\imath_{\text{S}}$ | Sending PID |
| | $e.\jmath_{\text{S}}$ | Recipient PID |
| | $e.\varsigma_{\text{S}}$ | *Undefined for send events* |
| ? | $e.\imath_{\text{S}}$ | Recipient PID |
| | $e.\jmath_{\text{S}}, e.\varsigma_{\text{S}}$ | *Undefined for receive events* |

| Index | Description |
|---|---|
| $m.\tau$ | Message type: event (evt) detach (dtc), routing (rtd) |
| $d.\imath_{\text{T}}$ | PID of tracer requesting detach of SuS PID $\imath_{\text{S}}$ |
| $d.\imath_{\text{S}}$ | PID of SuS process to stop tracing |
| $r.\imath_{\text{T}}$ | PID of tracer that starts routing message $m$ |
| $r.m$ | Embedded evt or dtc message being routed |

**(a)** Messages encoding *spawn*, *exit*, *send*, and *receive* events

**(b)** Detach and routing messages

**Table 1** Trace event (evt), detach request (dtc), and routing packet (rtd) message index names

| Requirement | Approach |
| --- | --- |
| $R_1$ Growing the set-up | Instrument tracers on-demand to create new trace partitions |
| $R_2$ Ensuring complete traces | Route trace events to deliver them to the correct tracer |
| $R_3$ Ensuring consistent traces | Prioritise routed trace events before others |
| $R_4$ Isolating tracers | Detach tracers from others once all trace events are routed |
| $R_5$ Minimising overhead | Target specific processes to instrument |
| $R_6$ Shrinking the set-up | Garbage collect redundant tracers and monitors |

**Table 2** RIARC approach to ensure trace soundness (def. 1) and reactive instrumentation (sec. 1)

## 3  Decentralised instrumentation

Our reason for encapsulating trace buffers and their coordination logic as tracers stems from the actor model. Trace buffers align with actor mailboxes, which localise the tracer state and enable tracers to run *independently*. The main logic replicated at each tracer is given in algs. 1–3. Tracers operate in two modes, *direct* (○) and *priority* (●), to counteract the effects of trace event reordering. We organise our tracer logic in algs. 1 and 3 to reflect these modes, respectively. Algs. 1 and 3 use the function AnalyseEvt, tasked with analysing events; see app. C.5.2 for details. Auxiliary tracer logic referenced in this section is relegated to app. A.

Every tracer maintains an internal state $\sigma$ consisting of the following three maps:

- the *routing* map, $\Pi$, governing how events are routed to other tracers,
- the *instrumentation* map, $\Lambda$, that determines which SuS processes to instrument, and
- the *traced-processes* map, $\Gamma$, tracking the SuS process set that the tracer currently traces.

Tbl. 2 summarises the challenges that RIARC needs to overcome to attain the reactive characteristics stated in sec. 1. Requirements $R_1$ and $R_6$ in tbl. 2 oblige the instrumentation to reorganise dynamically while the SuS executes to preserve its *elasticity*. Requirement $R_4$ offers a modicum of *resiliency* between the SuS and tracer processes, whereas $R_5$ minimises the instrumentation overhead by gathering only the events monitors require. This keeps the overall set-up *responsive*. Since RIARC builds on the actor model, it fulfils the *message-driven* requirement intrinsically. *Trace soundness* is safeguarded by requirements $R_2$ and $R_3$.

The operations Trace, Clear and Preempt give access to the tracing infrastructure. Trace($\iota_S,\iota_T$) enables a tracer with PID $\iota_T$ to register its interest in receiving trace events of a SuS process with PID $\iota_S$. This operation can be undone using Clear($\iota_S,\iota_T$), which *blocks* the calling tracer $\iota_T$ and returns once all the trace event messages for the SuS process $\iota_S$ that are in transit to the tracer $\iota_T$ have been delivered to $\iota_T$. It is worth remarking that this behaviour conforms to our proviso in sec. 1, *i.e.,* no communication faults. Preempt($\iota_S,\iota_T$) combines Clear and Trace. It enables the tracer pre-empting $\iota_T$ to take control of tracing the SuS process $\iota_S$ from another tracer $\iota'_T$ that is currently tracing $\iota_S$. Tracers use Clear or Preempt to modify the default process-tracing inheritance behaviour that tracing assumption $A_2$ describes. We refer to alg. 5 for the specifics of these operations.

We focus our presentation in secs. 3.1–3.6 of how RIARC addresses the challenges listed in tbl. 2 on the set-up of fig. 2b, where the processes $P$, $Q$ and $R$, are instrumented separately. This specific case highlights two aspects. First, it *emphasises* the complications that RIARC overcomes to establish the desired set-up while ensuring trace soundness, def. 1. Second, fig. 2b *covers all* other possible instrumentation set-ups. Disjoint sets of SuS processes, including the one shown in fig. 1c, can be obtained when tracers do not act on certain ⬦ (*spawn*) events, as sec. 3.1 explains. Notably, *any* centralised set-up, *e.g.* the one in fig. 1b, emerges naturally when the root tracer disregards all ⟡ events exhibited by the SuS.

▶ Note 6 (Naming conventions). For clarity, we adopt the convention that a SuS process $P$ is spawned from the signature $f_{s_P}$ and is assigned the PID $p_s$. A tracer for $P$ is named $T_P$ (short for $T_{\{P\}}$) and has the PID $p_T$. Other processes are treated likewise, *e.g.* the SuS process $Q$ has signature $f_{s_Q}$, PID $q_s$, while the tracer $T_Q$ for $Q$ has PID $q_T$, *etc.* ◀

## 3.1 Growing the set-up

Fig. 4 illustrates how the hierarchical creation sequence of SuS processes described in sec. 2.1 is exploited to instrument separate tracers. RIARC programs tracers to react to ⟡ (*spawn*) events in the trace. In fig. 4a, the root tracer $T_P$ traces process $P$, step ①. When $P$ spawns process $Q$, $Q$ automatically inherits $T_P$ (tracing assumption $A_2$ from sec. 2.1). Steps ② in fig. 4a emphasise that tracing inheritance is instantaneous. The event $e = \langle \text{evt}, \Diamond, p_s, q_s, f_{s_Q} \rangle$ is generated by $P$ when it spawns its child $Q$, step ③ in fig. 4a. The PID values of the parent and child processes carried by $e$, namely $p_s$ and $q_s$, are accessed via the indexes $e.\iota_s$ and $e.\jmath_s$ respectively (see tbl. 1a). Tracer $T_P$ uses this PID information to instrument a new tracer $T_Q$ for process $Q$ in step ④ of fig. 4b. By invoking PREEMPT($q_s, q_T$), $T_Q$ takes over tracing process $Q$ from the former tracer $T_P$ going forward. $T_Q$ creates a new trace partition for process $Q$ that is independent of the partition of $P$, step ⑤. Meanwhile, $T_P$ receives the send event $\langle \text{evt}, !, p_s, q_s \rangle$ in step ⑩ after $P$ messages $Q$ in step ⑥ of fig. 4c. Subsequent ⟡ events that $T_P$ or $T_Q$ may gather are handled as described in steps ③–⑤. Figs. 4c and 4d show how the final tracer $T_R$ is instrumented in step ⑫ after $Q$ spawns $R$ in step ⑧. As before, $T_Q$ traces $R$ automatically in step ⑧. $T_Q$ receives the event $\langle \text{evt}, \Diamond, q_s, r_s, f_{s_R} \rangle$ generated by $Q$ in step ⑪. $T_R$ invokes PREEMPT($r_s, r_T$) to create the trace partition for $R$ in step ⑬.

## 3.2 Ensuring complete traces

The asynchrony between the SuS and tracer processes can induce the interleaved execution shown in fig. 5, as an alternative execution to that shown in figs. 4b–4d. In fig. 5a, $T_P$ is slow

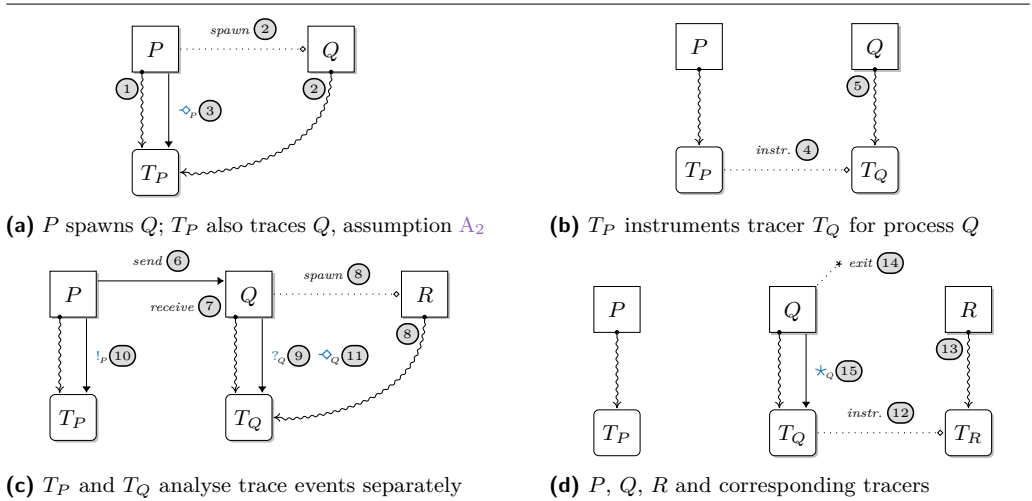

**(a)** $P$ spawns $Q$; $T_P$ also traces $Q$, assumption $A_2$    **(b)** $T_P$ instruments tracer $T_Q$ for process $Q$

**(c)** $T_P$ and $T_Q$ analyse trace events separately    **(d)** $P$, $Q$, $R$ and corresponding tracers

**Figure 4** Growing the tracer instrumentation set-up for processes $P$, $Q$ and $R$ (monitors omitted)

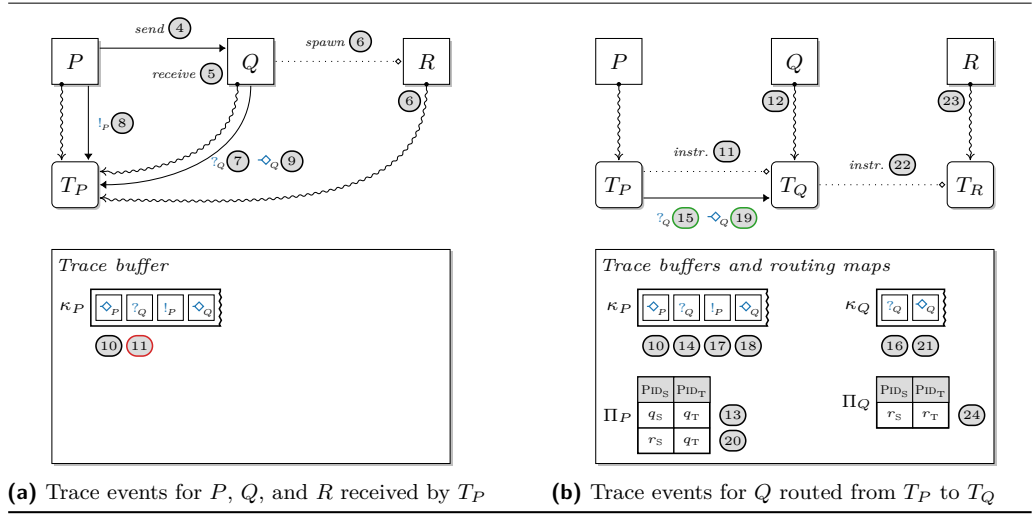

**(a)** Trace events for $P$, $Q$, and $R$ received by $T_P$    **(b)** Trace events for $Q$ routed from $T_P$ to $T_Q$

**Figure 5** Next-hop trace event routing using tracer routing maps $\Pi$ (monitors omitted)

to handle $\diamond_P$ it receives in ③ of fig. 4a and fails to instrument $T_Q$ promptly. Consequently, the events $?_Q$ and $\diamond_Q$ that $Q$ exhibits are sent to $T_P$ in steps ⑦ and ⑨ of fig. 5a. Step ⑪ shows the case where $\langle \text{evt},?,q_{\text{T}}\rangle$ is processed by $T_P$, rather than by the *intended* tracer $T_Q$ that would have been instrumented by $T_P$. This error breaches the *completeness* property of trace soundness w.r.t. $Q$, as the events $?_Q$ and $\diamond_Q$ meant for $Q$ reach the wrong tracer $T_P$.

To address this issue, RIARC uses a next-hop routing approach, where tracers *retain* the events they should handle and *forward* the rest to neighbouring tracers. We use the term *dispatch tracer* (*dispatcher* for short) to describe a tracer that receives trace events meant to be handled by another tracer. For instance, $T_P$ in fig. 5a becomes the dispatch tracer for $Q$ when it receives the events $?_Q$ and $\diamond_Q$ exhibited by $Q$, steps ⑦ and ⑨. We expect these events to be handled by $T_Q$ once it is instrumented. Dispatchers are tasked with embedding trace event (evt) or detach requests (dtc) into routing packet messages (rtd) and transmitting them to the next *known* hop. In fig. 5b, $T_P$ dispatches the events $?_Q$ and $\diamond_Q$ as follows. It first instruments $T_Q$ with $Q$ in step ⑪. Next, $T_P$ prepares $\langle \text{evt},?,r_{\text{S}}\rangle$ and $\langle \text{evt},\diamond,q_{\text{S}},r_{\text{S}},f_{\text{S}_R}\rangle$ for transmission by embedding each in rtd messages (steps ⑭ and ⑱). $T_P$ forwards the resulting routing packets, $\langle \text{rtd},\langle \text{evt},?,r_{\text{S}}\rangle\rangle$ and $\langle \text{rtd},\langle \text{evt},\diamond,q_{\text{S}},r_{\text{S}},f_{\text{S}_R}\rangle\rangle$, to its next-hop neighbour $T_Q$ in steps ⑮ and ⑲. The trace event $\langle \text{evt},!,p_{\text{S}},q_{\text{S}}\rangle$, however, is not forwarded but handled by $T_P$, as step ⑰ shows. Concurrently, $T_Q$ acts on the forwarded events $?_Q$ and $\diamond_Q$ in steps ⑯ and ㉑ and instruments $T_R$ as a result, step ㉒.

Tracers determine the events to retain or forward using the routing map, $\Pi\colon \text{PID}_{\text{S}} \rightharpoonup \text{PID}_{\text{T}}$. Every tracer queries its private routing map for each message it receives on SuS PID $m.\iota_{\text{S}}$. A tracer forwards a message to its neighbouring tracer with PID $\iota_{\text{T}}$ if a next-hop for that message exists, i.e., $\Pi(m.\iota_{\text{S}}) = \iota_{\text{T}}$. When the next-hop is undefined, i.e., $\Pi(m.\iota_{\text{S}}) = \bot$, $m$ is handled by the tracer. HANDLSPWN, HANDLEXIT and HANDLCOMM in alg. 1 implement this forwarding logic on lines 14, 23 and 31.

Dynamically populating the routing map is key to transmitting messages between tracers. A tracer adds the new mapping $e.\jmath_{\text{S}} \mapsto \jmath_{\text{T}}$ to its routing map $\Pi$ in case 1 or 2 below whenever it processes spawn trace events $e = \langle \text{evt},\diamond,\iota_{\text{S}},\jmath_{\text{S}},\varsigma_{\text{S}}\rangle$. One of two cases is considered for $e.\iota_{\text{S}}$:

1. $\Pi(\iota_{\text{S}}) = \bot$. The next-hop for $e$ is undefined, which cues the tracer to instrument the SuS process with PID $\jmath_{\text{S}}$. When applicable, the tracer processes the event *and* instruments a

**Algorithm 1** Logic handling ○ trace events, detach request dispatching, and forwarding

```
 1  def LOOP○(σ,ςM)                              35  def DISPATCHDTC(σ,d)
 2    forever do                                 36    match σ.Π(d.ιS) do
 3      m ← next message from trace buffer κ     37      case ⊥: fail dtc next-hop must be defined
 4      match m.τ do                             38      case JT:
 5        case evt: σ ← HANDLEVENT○(σ,ςM,m)     39        DISPATCH(d,JT)
 6        case dtc: σ ← DISPATCHDTC○(σ,ςM,m)              # Next-hop for d.ιS no longer needed
 7        case rtd: σ ← FORWDRTD○(σ,ςM,m)       40        σ.Π ← σ.Π\{⟨d.ιS,JT⟩}
                                                 41        TRYGC(σ)
 8  def HANDLEVT○(σ,ςM,e)                        42    return σ
 9    match e.ℓ do
10      case ⋄: return HANDLSPWN○(σ,ςM,e)      43  def FORWDRTD○(σ,r)
11      case ⋆: return HANDLEXIT○(σ,ςM,e)      44    m ← r.m  # Read embedded message in r
12      case !,?: return HANDLCOMM○(σ,ςM,e)    45    match m.τ do
                                                 46      case dtc: return FORWDDTC(σ,r)
13  def HANDLSPWN○(σ,ςM,e)                       47      case evt: return FORWDEVT(σ,r)
14    match σ.Π(e.ιS) do
15      case ⊥: # No next-hop for e.ιS; handle e 48  def FORWDDTC(σ,r)
16        ANALYSEEVT(ςM,e) # App. C.5.2         49    d ← r.m
17        σ ← INSTRUMENT○(σ,e,self())          50    match σ.Π(d.ιS) do
18      case JT: # Next-hop for e.ιS exists via JT 51    case ⊥: fail dtc next-hop must be defined
19        DISPATCH(e,JT)                         52      case JT:
          # Set next-hop of e.JS to tracer of e.ιS 53      FORWD(r,JT)
20        σ.Π ← σ.Π∪{⟨e.JS,JT⟩}                         # Next-hop for d.ιS no longer needed
21    return σ                                   54        σ.Π ← σ.Π\{⟨d.ιS,JT⟩}
                                                 55        TRYGC(σ)
22  def HANDLEXIT○(σ,ςM,e)                       56    return σ
23    match σ.Π(e.ιS) do
24      case ⊥: # No next-hop for e.ιS; handle e 57  def FORWDEVT(σ,r)
25        ANALYSEEVT(ςM,e) # App. C.5.2         58    e ← r.m
26        σ.Γ ← σ.Γ\{⟨e.ιS,○⟩}                  59    match σ.Π(e.ιS) do
27        TRYGC(σ)                               60      case ⊥: fail evt next-hop must be defined
28      case JT: DISPATCH(e,JT)                  61      case JT:
29    return σ                                   62        FORWD(r,JT)
                                                          # For spawn events, tracer also sets a
30  def HANDLCOMM○(σ,ςM,e)                               # new next-hop for e.JS
31    match σ.Π(e.ιS) do                                 # Next-hop of e.JS to same tracer of e.ιS
32      case ⊥: ANALYSEEVT(ςM,e) # App. C.5.2   63        if (e.ℓ = ⋄)
33      case JT: DISPATCH(e,JT)                  64          σ.Π ← σ.Π∪{⟨e.JS,JT⟩}
34    return σ                                   65    return σ
```

separate tracer with PID $J_T$. It then adds the mapping $e.J_S \mapsto J_T$ to $\Pi$. The tracer leaves $\Pi$ *unmodified* and handles the event itself if a separate tracer is not required. Opting for a separate tracer is determined by the instrumentation map $\Lambda$, as discussed in sec. 3.5.

**2.** $\Pi(\iota_S) = J_T$. The next-hop for $e$ is defined, and the tracer forwards the event to the neighbouring tracer $J_T$. The tracer also records a new next-hop by adding $e.J_S \mapsto J_T$ to $\Pi$.

The addition of $e.J_S \mapsto J_T$ in cases 1 and 2 ensures that future events originating from $J_S$ can always be forwarded via a next-hop to a neighbouring tracer $J_T$ (see invariants on lines 37, 51, and 60). Fig. 5b shows the routing maps of the tracers $T_P$ and $T_Q$. $T_P$ adds $q_S \mapsto q_T$ in step ⑬ after processing $\langle \text{evt}, \diamond, p_S, q_S, f_{S_Q} \rangle$ from its trace buffer in ⑩. $T_P$ then instruments $Q$ with the tracer $T_Q$ in step ⑪; an instance of case 1. The function INSTRUMENT in alg. 2 details this on line 4, where the mapping $e.J_S \mapsto J_T$ is added to $\Pi$ following the creation of

**Algorithm 2** Tracer instrumentation operations for direct (○) and priority (●) modes

**Expect:** $e = \langle \text{evt}, \diamond, \iota_{\text{S}}, \jmath_{\text{S}}, \varsigma_{\text{S}} \rangle$      **Expect:** $e = \langle \text{evt}, \diamond, \iota_{\text{S}}, \jmath_{\text{S}}, \varsigma_{\text{S}} \rangle$

1   **def** $\text{INSTRUMENT}_\circ(\sigma, e, \iota_{\text{T}})$
2     **if** $((\varsigma_{\text{M}} \leftarrow \sigma.\Lambda(e.\varsigma_{\text{S}})) \neq \bot)$
        *# New tracer $\jmath_T$ for new SuS process $e.\jmath_S$*
3       $\jmath_{\text{T}} \leftarrow \textsf{spwn}(\text{TRACER}(\sigma, \varsigma_{\text{M}}, e.\jmath_{\text{S}}, \iota_{\text{T}}))$
4       $\sigma.\Pi \leftarrow \sigma.\Pi \cup \{\langle e.\jmath_{\text{S}}, \jmath_{\text{T}} \rangle\}$
5     **else**
        *# In ○ mode, this tracer has detached*
        *# all processes from its dispatcher $\iota_T$*
        *# This tracer traces new SuS process $e.\jmath_S$*
        *# by tracing inheritance assumption $A_2$*
6       $\sigma.\Gamma \leftarrow \sigma.\Gamma \cup \{\langle e.\jmath_{\text{S}}, \circ \rangle\}$
7     **return** $\sigma$

8   **def** $\text{INSTRUMENT}_\bullet(\sigma, e, \iota_{\text{T}})$
9     **if** $((\varsigma_{\text{M}} \leftarrow \sigma.\Lambda(e.\varsigma_{\text{S}})) \neq \bot)$
        *# New tracer $\jmath_T$ for new SuS process $e.\jmath_S$*
10     $\jmath_{\text{T}} \leftarrow \textsf{spwn}(\text{TRACER}(\sigma, \varsigma_{\text{M}}, e.\jmath_{\text{S}}, \iota_{\text{T}}))$
11     $\sigma.\Pi \leftarrow \sigma.\Pi \cup \{\langle e.\jmath_{\text{S}}, \jmath_{\text{T}} \rangle\}$
12   **else**
        *# In ● mode, this tracer must detach*
        *# SuS process $e.\jmath_S$ from its dispatcher $\iota_T$*
13     $\text{DETACH}(e.\jmath_{\text{S}}, \iota_{\text{T}})$
        *# This tracer traces new SuS process $e.\jmath_S$*
14     $\sigma.\Gamma \leftarrow \sigma.\Gamma \cup \{\langle e.\jmath_{\text{S}}, \bullet \rangle\}$
15   **return** $\sigma$

tracer $\jmath_{\text{T}}$, line 3. Step ⑳ of fig. 5b is an instance of case 2. Here, $T_P$ adds $r_{\text{S}} \mapsto q_{\text{T}}$ to $\Pi_P$ after processing $\langle \text{evt}, \diamond, q_{\text{S}}, r_{\text{S}}, f_{\text{S}_R} \rangle$ for $R$ in step ⑱ since $\Pi_P(q_{\text{S}}) = q_{\text{T}}$. Crucially, $T_P$ *does not* instrument a new tracer, but delegates the task to $T_Q$ by forwarding $\diamond_Q$. Lines 20 and 64 in alg. 1 (and later line 24 in alg. 3) are manifestations of this, where the mapping $e.\jmath_{\text{S}} \mapsto \jmath_{\text{T}}$ is added after the $\diamond$ event $e$ is forwarded to the next-hop $\jmath_{\text{T}}$. $T_Q$ instruments the SuS process $R$ in step ㉒ with $T_R$, which has the PID $r_{\text{T}}$. It then adds the mapping $r_{\text{S}} \mapsto r_{\text{T}}$ to $\Pi_Q$ in step ㉔, as no next-hop is defined for $q_{\text{S}}$, *i.e.,* $\Pi_Q(q_{\text{S}}) = \bot$. Henceforth, any events exhibited by $R$ and received at $T_P$ can be dispatched by the latter tracer through $T_Q$ to $T_R$.

We note that every tracer is only aware of its neighbouring tracers. This means messages may pass through multiple tracers before reaching their intended destination. Next-hop routing keeps the logic inside RIARC straightforward since tracers forward messages based solely on local information in their routing map. Such an approach makes the instrumentation set-up readily adaptable to dynamic changes in the SuS, is easier to scale, and has been shown to induce lower latency when compared to general routing strategies [83, 107]. The DAG of interconnected tracers induced by next-hop routing ensures that every message is eventually delivered to the correct tracer if a path exists or is handled by the tracer otherwise. Fig. 5b illustrates this concept, where the next-hop mappings inside $\Pi_P$ point to $T_Q$, and the mappings in $\Pi_Q$ point to $T_R$ in turn. Consequently, any events that $R$ exhibits and that $T_P$ receives are forwarded *twice* to reach the target tracer $T_R$: from tracer $T_P$ to $T_Q$, and from $T_Q$ to $T_R$. RIARC relies on the operations DISPATCH and FORWD to accomplish next-hop routing (see alg. 4 in app. A). DISPATCH creates a routing packet $\langle \iota_{\text{S}}, m \rangle$ and embeds the trace event or detach message $m$ to be routed. Alg. 1 shows how routing packets are handled by tracers. For instance, FORWDEVT extracts the embedded message from the routing packet on line 58 and queries the routing map to determine the next-hop, line 59. If it does, the packet is forwarded, as $\text{FORWD}(r, \jmath_{\text{T}})$ on line 62 indicates. Crucially, the **fail** invariant on line 60 asserts that the next-hop for a routing packet is *always* defined. The cases for DISPATCHDTC and FORWDDTC in alg. 1 are analogous.

### 3.3   Ensuring consistent traces

Next-hop routing alone does not guarantee trace consistency, *i.e.,* that the order of events in the trace reflects the one in which these occur locally at SuS processes, def. 1. Trace event reordering arises when a tracer gathers events of a SuS process (we call these *direct events*) and simultaneously receives *routed events* concerning said process from other tracers.

Fig. 6a gives another interleaving to the one of fig. 5b to underscore the deleterious effect such a race condition provokes when events are reordered at $T_Q$. In step ⑫ $T_Q$ takes over $T_P$ to continue tracing process $Q$. $T_Q$ collects the event $\star_Q$ in step ⑮, which happens before $T_Q$ receives the routed event $?_Q$ concerning $Q$ in step ⑰ of fig. 6a. If $T_Q$ processes events from its trace buffer $\kappa_Q$ in sequence, as in step ⑱, it violates trace consistency w.r.t. $Q$ (the correct trace ordering should be '$?_Q.\diamond_Q.\star_Q$'). Naïvely handling $\star$ before $?$ erroneously reflects that $Q$ receives messages after it terminates.

RIARC tracers resolve this issue by prioritising the processing of routed trace events using selective message reception [42]. In doing so, tracers encode the invariant that '*routed* events temporally precede all others that are gathered *directly* by the tracer'. RIARC tracers operate in one of two modes, priority (●) and direct (○), which adequately distinguishes past (*i.e.,* routed) and current (*i.e.,* direct) events from the perspective of the tracer receiving them.

Fig. 6b illustrates this concept. It shows that when in priority mode, $T_Q$ dequeues the routed events $?_Q$ and $\diamond_Q$ labelled by ● first. The event $?_Q$ is handled in step ㉓, whereas $\diamond_Q$ results in the instrumentation of tracer $T_R$ in step ㉕ of fig. 6b. Meanwhile, $T_Q$ can still receive events directly from $Q$ while priority events are being handled. Yet, direct trace events from $Q$ are considered only *after* $T_Q$ transitions to direct mode. Newly-instrumented tracers default to ● mode to implement the described logic; see line 14 in alg. 4 of app. A.

Loop● in alg. 3 shows the logic prioritising routed events, which are dequeued on line 3 and handled on line 6. HandlSpwn, HandlExit, and HandlComm in Loop○ and Loop● handle events *differently*. A tracer in direct mode performs *one* of three actions (see alg. 1):

1. it *analyses* events for RV purposes via the function AnalyseEvt($\varsigma_M$,$e$), *e.g.* line 32,
2. it *dispatches* events that it directly gathers using Dispatch($e$,$\jmath_T$), when events ought to be handled by other tracers, *e.g.* line 33, or
3. it *forwards* routed events to the next-hop through Forwd($r$,$\jmath_T$), *e.g.* line 62.

Tracers in priority mode exclusively handle routed messages as points 1 and 3 describe, *e.g.* lines 38 and 39 in alg. 3. However, no event dispatching is performed.

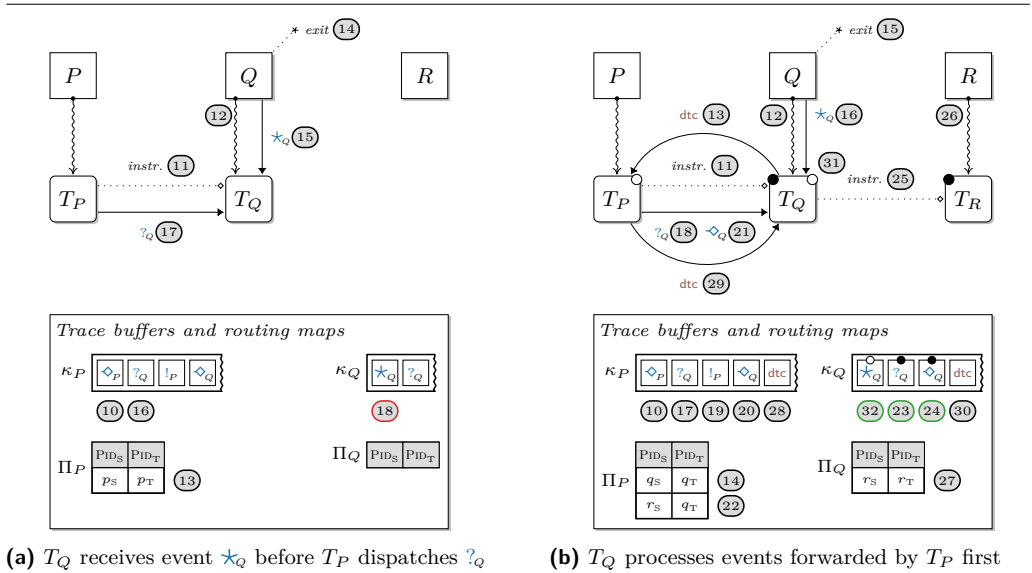

**(a)** $T_Q$ receives event $\star_Q$ before $T_P$ dispatches $?_Q$  **(b)** $T_Q$ processes events forwarded by $T_P$ first

**Figure 6** Trace event reordering using priority (●) and direct (○) tracer modes (monitors omitted)

**Algorithm 3** Logic handling ● trace events, detach request acknowledgements, and forwarding

```
 1  def LOOP●(σ,ςM)                           26  def HANDLEXIT●(σ,ςM,r)
 2   forever do                               27   e ← r.m
 3    r ← next rtd message from trace buffer κ 28   match σ.Π(e.ιS) do
 4    m ← r.m  # Read embedded message in r    29    case ⊥: # No next-hop for e.ιS; handle e
 5    match m.τ do                             30     ANALYSEEVT(ςM,e) # App. C.5.2
 6     case evt: σ ← HANDLEVT●(σ,ςM,r)          31     σ.Γ ← σ.Γ\{⟨e.ιS,●⟩}
 7     case dtc:                               32     TRYGC(σ)
          # dtc ack relayed from dispatch tracer 33    case JT: FORWD(r,JT)
 8      σ ← HANDLDTC(σ,ςM,r)                   34   return σ

 9  def HANDLEVT●(σ,ςM,r)                     35  def HANDLCOMM●(σ,ςM,r)
10   e ← r.m                                  36   e ← r.m
11   match e.ℓ do                             37   match σ.Π(e.ιS) do
12    case ◇: return HANDLSPWN●(σ,ςM,r)        38    case ⊥: ANALYSEEVT(ςM,e) # App. C.5.2
13    case ⋆: return HANDLEXIT●(σ,ςM,r)         39    case JT: FORWD(r,JT)
14    case !,?: return HANDLCOMM●(σ,ςM,r)      40   return σ

15  def HANDLSPWN●(σ,ςM,r)                    41  def HANDLDTC(σ,ςM,r)
16   e ← r.m                                  42   d ← r.m
17   match σ.Π(e.ιS) do                       43   match σ.Π(d.JS) do
18    case ⊥: # No next-hop for e.ιS; handle e  44    case ⊥:
19     ANALYSEEVT(ςM,e) # App. C.5.2           45     assert d.ιT = self()  unexpected dtc ack
20     ιT ← r.ιT # Read PID of dispatch tracer  46     σ.Γ ← (σ.Γ\{⟨d.JS,●⟩}) ∪ {⟨d.JS,○⟩}
21     σ ← INSTRUMENT●(σ,e,ιT)                 47     if ({⟨ιS,γ⟩ | ⟨ιS,γ⟩ ∈ σ.Γ, γ = ●} = ∅)
22    case JT: # Next-hop for e.ιS exists via JT 48      LOOP○(σ,ςM) # Put tracer in ○ mode
23     FORWD(r,JT)                             49    case JT:
       # Set next-hop of e.JS to tracer of e.ιS  50    assert d.ιT ≠ self()  dtc meant for ιT
24     σ.Π ← σ.Π ∪ {⟨e.JS,JT⟩}                 51     FORWD(r,JT)
25   return σ                                 52   return σ
```

## 3.4 Isolating tracers

A tracer in priority mode coordinates with the dispatch tracer of a particular SuS process it traces. This enables the tracer to determine when *all* of the events of that process have been routed to it by the dispatch tracer. The negotiation is effected using dtc, which the tracer sends to the relevant dispatch tracer. Each tracer records the set of processes it traces in the *traced-processes map*, $\Gamma : \text{PID}_S \rightharpoonup \{○,●\}$. A SuS process mapping is added to $\Gamma$ when a tracer starts gathering trace events for that process and removed once the process terminates. Lines 6 and 14 in alg. 2 add fresh mappings to $\Gamma$; lines 26 in alg. 1 and 31 in alg. 3 purge mappings from $\Gamma$. A tracer in priority mode must issue a dtc request *for each* process it tracks in $\Gamma$ before it can transition to direct mode and start operating on the trace events it gathers directly. The detach request, $d = \langle \text{dtc}, \iota_T, \iota_S \rangle$, contains the PIDs of the issuing tracer and the SuS process to be detached from the dispatch tracer. Once the tracer receives an acknowledgement to the dtc request for the SuS PID $d.\iota_S$ from the dispatch tracer, it updates the corresponding entry $d.\iota_S \mapsto ●$ in $\Gamma$, marking it as detached, $d.\iota_S \mapsto ○$. Alg. 3 shows this logic on line 46. A tracer transitions from priority to direct mode once *all* the processes in its $\Gamma$ map are marked detached; line 47 in alg. 3 performs this check. Once in direct mode, tracers are isolated from others in the choreography.

Fig. 6b depicts the tracer $T_Q$ in priority mode sending the detach request $\langle \text{dtc}, q_T, q_S \rangle$ for SuS PID $q_S$ to the dispatch tracer. This happens in step ⑬, after $T_Q$ starts tracing $Q$ directly in step ⑫. Alg. 2 effects this transaction with the dispatch tracer by the operation DETACH on line 13; see app. A for definition of DETACH. The dtc request issued by $T_Q$ is deposited in the trace buffer of the dispatch tracer $T_P$ after the events $?_Q$ and $\diamond_Q$. $T_P$ processes the messages in its buffer sequentially in ⑩, ⑰, ⑲, ⑳ and ㉘, and forwards $?_Q$ and $\diamond_Q$ to $T_Q$, steps ⑱ and ㉑. Crucially, $T_P$ *acknowledges* the dtc request issued by $T_Q$: $T_P$ dispatches dtc back to tracer $T_Q$, as step ㉙ indicates. $T_Q$ first handles the events $?_Q$ and $\diamond_Q$ (tagged with • in fig. 6b) in steps ㉓ and ㉔. Lastly, $T_Q$ handles dtc in ㉚ and marks process $Q$ as detached from its dispatch tracer $T_P$. The update on the traced-process map $\Gamma$ is performed by HANDLDTC on line 46 in alg. 3. Tracer $T_Q$ in fig. 6b transitions to direct mode in step ㉛, when the only process $Q$ that it traces is detached. $T_Q$ resumes handling $\star_Q$ in step ㉜, which is consistent w.r.t. the events exhibited locally at $Q$, *i.e.*, '$?_Q.\diamond_Q.\star_Q$'.

An acknowledgement to a detach request sent from a dispatch tracer, $\langle \mathsf{dtc}, \iota_\mathrm{T}, \iota_\mathrm{S} \rangle$, is generally propagated through multiple next-hops before it reaches the tracer with PID $\iota_\mathrm{T}$ issuing the request. Since a dtc request informs the dispatch tracer that $\iota_\mathrm{T}$ is gathering trace events for the SuS PID $\iota_\mathrm{S}$ *directly*, the next-hop entries in the routing maps of tracers on the DAG path from the dispatch tracer to $\iota_\mathrm{T}$ are *stale*. Each tracer on this DAG path purges the next-hop entry for the SuS PID $\iota_\mathrm{S}$ in $\Gamma$ once it forwards dtc to the neighbouring tracer. DISPATCHDTC and FORWDDTC in alg. 1 perform this clean-up. Fig. 6b does not illustrate the latter clean-up flow, which we summarise next. After receiving dtc, the dispatch tracer $T_P$ removes from $\Pi_P$ the next-hop mapping $q_\mathrm{S} \mapsto q_\mathrm{T}$ and calls DISPATCHDTC to acknowledge the detach request $\langle \mathsf{dtc}, q_\mathrm{T}, q_\mathrm{S} \rangle$ it receives from $T_Q$. Similarly, $T_P$ removes $r_\mathrm{S} \mapsto q_\mathrm{T}$ once it acknowledges the detach request $\langle \mathsf{dtc}, r_\mathrm{T}, r_\mathrm{S} \rangle$ sent from $T_R$. Once $T_Q$ receives the routing packet $\langle \mathsf{rtd}, p_\mathrm{T}, \langle \mathsf{dtc}, r_\mathrm{T}, r_\mathrm{S} \rangle \rangle$ that embeds the detach acknowledgement $T_P$ sends, it removes the next-hop mapping $r_\mathrm{S} \mapsto r_\mathrm{T}$ from $\Pi_Q$. $T_Q$ then forwards this dtc acknowledgement to $T_R$.

RIARC ensures that all routing packets carrying dtc acknowledgements terminate at the tracers that issued these dtc requests. This requires *one* of two tracer conditions to hold:

**1.** either the tracer cannot forward the dtc acknowledgement to a next-hop, meaning that the tracer sent the dtc request, or

**2.** the tracer can forward the dtc acknowledgement via a next-hop, in which case the tracer did not issue the dtc request.

Alg. 3 enforces this invariant on lines 44 and 45 for case 1, and on lines 49 and 50 for case 2.

### 3.5 Minimising overhead

Instrumenting specific processes—in contrast to fully instrumenting the SuS—reduces the volume of gathered trace events and helps lower the runtime overhead induced. RIARC uses the instrumentation map, $\Lambda:\mathrm{SIG_S} \rightharpoonup \mathrm{SIG_M}$, to this end. $\Lambda$ specifies the SuS function signatures to instrument and the corresponding RV monitor signatures tasked with the analysis via ANALYSEEVT. RIARC utilises the signature $e.\varsigma_\mathrm{S}$ carried by spawn events $e = \langle \mathsf{evt}, \diamond, \iota_\mathrm{S}, \jmath_\mathrm{S}, \varsigma_\mathrm{S} \rangle$ to determine whether the SuS process spawning $e.\varsigma_\mathrm{S}$ requires a separate tracer. The INSTRUMENT operations in alg. 2 perform this check against $\Lambda$ (lines 2 and 9). If a separate tracer is not required, $e.\jmath_\mathrm{S}$ is instrumented using the tracer of its parent process, $e.\iota_\mathrm{S}$; see tracing assumptions $A_1$ and $A_2$. This logic caters for all the set-ups shown in figs. 1b, 1c, and 2b.

### 3.6 Shrinking the set-up

RIARC remains elastic by discarding unneeded tracers. Tracers in direct and priority mode purge SuS PID references from the traced-process map when handling $\star$ trace events. HANDLEXIT$_\circ$ and HANDLEXIT$_\bullet$ implement this logic in algs. 1 and 3 on lines 26 and 31.

Tracer termination does *not* occur when the tracer has no processes left to trace, *i.e.,* when $\Gamma = \emptyset$, since the tracer may be required to forward trace events to neighbouring tracers. Instead, tracers perform a garbage collection check each time a mapping from $\Gamma$ or $\Pi$ is removed. A tracer terminates when $\Gamma = \Pi = \emptyset$, indicating that it has no SuS processes left to trace or any next-hop forwarding to perform. TRYGC used on lines 27, 41, and 55 in alg. 1, as well as on line 32 in alg. 3 encapsulates this check. Note that garbage collection never prematurely disrupts the RV analysis that tracers conduct, as invocations to ANALYSEEVT always precede TRYGC checks in our logic of algs. 1 and 3.

## 4    Correctness validation

We assess the validity of RIARC in two stages. First, we confirm its implementability by instantiating the core logic of algs. 1–3 to Erlang. Our implementation targets two RV scenarios: online and offline monitoring [65, 21]. Second, we subject the implementation to a series of systematic tests using a selection of instrumentation set-ups. These tests exhaustively emulate the interleaved execution of the SuS and tracer processes by generating all the *valid* permutations of events in a set of traces. This exercises the tracer choreography invariants mentioned in sec. 3, confirming the integrity of the tracer DAG topology under each interleaving. We also use specialised RV monitor signatures in ANALYSEEVT to assert the soundness (def. 1) of trace event sequences analysed by tracers; see algs. 1 and 3 in sec. 3.

### 4.1    Implementability

Our implementation of RIARC maps the tracer processes from sec. 3 to Erlang actors[2]. The routing ($\Pi$), instrumentation ($\Lambda$), and traced-processes ($\Gamma$) maps constituting the tracer state $\sigma$ are realised as Erlang maps for efficient access. Trace event buffers $\kappa$ coincide with actor mailboxes, while the remaining logic in algs. 1–3 translates directly to Erlang code. This one-to-one mapping gives us confidence that our implementation reflects the algorithm logic.

In *online* RV, monitors analyse trace events while the SuS executes, whereas the *offline* setting defers this analysis until the system terminates. Fig. 11 in app. B.1 captures the distinction in process tracing between online and offline instrumentation in our setting (showing trace buffers only). The online instrumentation set-up (fig. 11a) employs the tracing infrastructure offered by the EVM, which deposits SuS trace event messages in tracer mailboxes. Erlang tracing complies with tracing assumption $A_1$, enabling RIARC to instrument disjoint SuS processes sets. We configure the EVM with the `set_on_spawn` flag so that spawned processes automatically inherit the same tracer as their parent [42]. This tracer assignment is atomic, meeting tracing assumption $A_2$. We also use the `procs`, `send`, and `receive` tracing flags, which constrain the events emitted by the EVM to ⟡, ⋆, !, and ⋆. The EVM enforces single-process tracing, *i.e.,* tracing assumption $A_3$, and guarantees that ⟡ events of descendant processes are causally-ordered [137], *i.e.,* tracing assumption $A_4$.

The offline counterpart differs only in its tracing layer, where events are read as *recorded* runs of the SuS. Recorded runs can be obtained externally, *e.g.* using DTrace [37] or LTTng [56], making it possible to monitor systems that execute outside of the EVM. Our bespoke offline tracing engine of fig. 11b fulfils tracing assumptions $A_1$–$A_4$. This is crucial since it permits the *same* implementation of RIARC to be used in online and offline settings. Sec. 4.2 leverages this aspect to validate RIARC exhaustively using trace permutations.

---

[2] The artefact may be found at https://doi.org/10.5281/zenodo.10634182.

We develop two versions of the TRACE, CLEAR, and PREEMPT functions of alg. 5 to standardise the tracing API for online and offline use. The overloads for online use give access to the EVM tracing via the Erlang built-in primitive `trace` [42]. The second set of overloads wraps around our offline tracing engine to replay files containing specifically-formatted trace events. Offline tracing relaxes tracing assumption $A_4$, as recorded runs do not generally guarantee that the ⬦ events of descendant SuS processes are causally ordered. Our offline tracing logic relies on the PID information carried by ⬦ events to rearrange them causally and recover the causal ordering per tracing assumption $A_4$. TRACE($\iota_S, \iota_T$) registers a tracer $\iota_T$ with the offline tracing engine, which maintains an event buffer for $\iota_T$, together with a set of SuS PIDs that $\iota_T$ traces. A tracer can use TRACE with multiple SuS PIDs to register to obtain events for a set of processes, *i.e.,* tracing assumption $A_1$. The tracing engine accumulates the events it reads from file in each tracer buffer and delivers events to the corresponding tracer mailbox once the casual ordering between ⬦ events of descendant SuS processes is established. Our offline tracing engine implements tracing inheritance (tracing assumption $A_2$) and enforces single-process tracing (tracing assumption $A_3$). Ex. 7 in app. B.1 sketches how the tracing engine uses its internal tracer buffers to deliver events to tracers.

## 4.2 Correctness

Conventional testing does not guarantee the absence of concurrency errors due to the different interleaved executions that may be possible [108]. While subjecting the system under test to high loads raises the likelyhood of obtaining more coverage, this still depends on external factors, such as scheduling, which dictate the executions induced in practice. Controlling the conditions for concurrency testing requires a *systematic exploration* of all the interleaved executions [77]. In fact, it is *not the size* of the testing load that matters, but the choice of interleaved executions that exhaust the space of possible system states [13]. Concuerror [48] is a tool for systematic Erlang code testing. Unfortunately, we could not use Concuerror to test our RIARC implementation, as we were unable to integrate it with Erlang tracing.

We, nevertheless, adopt the systematic scheme advocated by Concuerror. Our approach uses the offline tracing tool described in sec. 4.1 to induce specific interleaved sequences for instrumentation set-ups, such as those of figs. 1b, 1c, and 2a. We obtain these sequences by taking all the sound (def. 1) event permutations of traces produced by the SuS. These sequences are then replayed by the offline tracing engine to systematically induce interleaving sequences in the SuS. Our final RIARC implementation embeds additional invariants besides the ones mentioned in sec. 3, *e.g.* the **assert** and **fail** statements in algs. 1 and 3. Readers are referred to app. B.2 for the full list. We ascertain *trace soundness* for each SuS interleaving that is emulated. This is accomplished via the function ANALYSEEVT, which we preload with monitors that assert the event sequence expected at each tracer. We also use identical tests in our empirical evaluation of sec. 5 under high loads. It is worth mentioning that while we systematically drive the execution of the SuS, we do not control the execution of tracers. Yet, we indirectly induce various dynamic tracer arrangements in the monitor DAG topology under the different groupings of SuS process sets that tracers instrument. For example, we fully instrument system depicted in fig. 2a in all its configurations, *e.g.* $\mathcal{C}_1 = [T_{\{P\}} \rightsquigarrow \{P\}, T_{\{Q\}} \rightsquigarrow \{Q\}, T_{\{R\}} \rightsquigarrow \{R\}]$, $\mathcal{C}_2 = [T_{\{P,Q\}} \rightsquigarrow \{P,Q\}, T_{\{R\}} \rightsquigarrow \{R\}]$, ..., $\mathcal{C}_5 = [T_{\{P,Q,R\}} \rightsquigarrow \{P,Q,R\}]$, as well as instrument it partially, *e.g.* $\mathcal{C}_6 = [T_{\{P\}} \rightsquigarrow \{P\}]$, $\mathcal{C}_7 = [T_{\{P,Q\}} \rightsquigarrow \{P,Q\}]$, *etc.* Each of these configurations, when individually paired with every fabricated interleaved execution of the SuS, indicate that our RIARC implementation and corresponding logic of sec. 3 is correct.

## 5 Empirical evaluation

We assess the feasibility of our RIARC implementation, confirming it safeguards the *responsive*, *resilient*, *message-driven*, and *elastic* attributes of the SuS. Sec. 4 targets a small selection of instrumentation set-ups to induce interleaved execution sequences and validate correctness exhaustively. We now employ *stress testing* [112] to investigate how RIARC performs in terms of the *runtime overhead* it exhibits. Our study focusses on *online* monitoring, as its overhead requirement is far more stringent than offline monitoring [64, 65, 21, 74]. We evaluate RIARC against inline instrumentation since the latter is regarded as the most efficient instrumentation technique [63, 62, 21]. This comparison establishes a solid basis for our results to be generalised reliably. We also compare RIARC to centralised instrumentation to confirm that the latter approach does not scale under typical loads.

Our experiments are extensive. We use two hardware platforms to model edge-case scenarios based on limited hardware and general-case scenarios using commodity hardware. The evaluation subjects inline, centralised, and RIARC instrumentation to high loads that go beyond the state of the art and use realistic workload profiles. We gauge overhead under three performance metrics, the *response time*, *memory consumption*, and *scheduler utilisation*, which are crucial for reactive systems [7, 112]. Our results confirm that the overhead RIARC induces is adequate for applications such as soft real-time systems [42, 97], where the latency requirement is typically in the order of seconds [95]. We also show that RIARC yields overhead comparable to inlining in settings exhibiting moderate concurrency.

### 5.1 Benchmarking tool

Benchmarking is standard practice for gauging runtime overhead in software [103, 80, 36]. Frameworks, including DaCapo [28] and Savina [87], offer limited concurrency, making them inapplicable to our case; see App. C.1 for detailed reasons. Industry-proven *synthetic* load testing benchmarking tools cater to reactive systems, *e.g.* Apache JMeter [70], Tsung [118], and Basho Bench [23]. Their general-purpose design, however, necessarily treats systems as a black box by gathering metrics externally, which may impact measurement *precision* [7]. Moreover, these load testers generate standard workloads, *e.g.* Poisson processes [82, 105, 92], but lack others, *e.g.* load bursts, that replicate typical operation or induce edge-case stress.

We adopt BenchCRV [7], another synthetic load tester specific to RV benchmarking for reactive systems. It sets itself apart from the tools above because it does not require external software (*e.g.*, a web server) to drive tests. Instead, BenchCRV produces different models that *closely emulate* real-world software behaviour. These models are based on the master-worker paradigm [127]: a pervasive architecture in distributed (*e.g.* Big Data frameworks, render farms) and concurrent systems [138, 76, 55, 141]. Like Tsung and Basho Bench, BenchCRV exploits the lightweight EVM process model to generate highly-concurrent workloads.

BenchCRV creates master-worker models and induces workloads derived from configurable parameters. In these models, the master process spawns a series of workers and allocates tasks. The volume of workers per benchmark run is set via the parameter $n$. Each worker task consists of a *batch* of requests that the worker receives, processes, and echoes back to the master process. The amount of requests batched in one task is given by the parameter $w$. Workers terminate when all of their allotted tasks are processed and acknowledged by the master. BenchCRV creates workers based on *workload profiles*. A profile dictates how the master spreads its creation of workers along the loading timeline, $t$, given in seconds. BenchCRV supports three workload profiles based on ones typical in practice (*e.g.* see fig. 13):
**Steady** models the SuS under stable workload (Poisson process).

**Pulse** models the SuS under gradually rising and falling workload (Normal distribution).
**Burst** models the SuS under stress due to workload spikes (Log-normal distribution).
The tool records three performance metrics to give a multi-faceted view of system overhead:
**Mean response time** in milliseconds (ms), gauging monitoring latency effects on the SuS.
**Mean memory consumption** in GB, gauging monitoring memory pressure on the SuS.
**Mean scheduler utilisation** as a percentage of the total processing capacity, showing how monitors maximise the scheduler use.

The prevalent use of the master-worker paradigm, the veracity with which BenchCRV models systems, the range of realistic workload profiles, and the choice of runtime metrics it gathers make this tool ideal for our experiments. Readers are referred to app. C.2 and [7] for details.

## 5.2 Benchmark configuration

The BenchCRV master-worker models we generate take the role of the SuS in our experiments. We consider *edge-case* and *general-case* hardware platform set-ups for the following reasons:

**$P_E$ Edge-case** captures platforms with *limited* hardware. It uses an Intel Core i7 M620 64-bit CPU with 8GB of memory, running Ubuntu 18.04 LTS and Erlang/OTP 22.2.1.

**$P_G$ General-case** captures platforms with *commodity* hardware. It uses an Intel Core i9 9880H 64-bit CPU with 16GB of memory, running macOS 12.3.1 and Erlang/OTP 25.0.3.

The EVMs on platforms $P_E$ and $P_G$ are set with 4 and 16 scheduling threads, respectively. These scheduler settings coincide with the processors available on each SMP [11] platform. We also use the $P_E$ and $P_G$ platforms with two concurrency scenarios for reactive systems:

**$C_H$ High concurrency scenarios** perform short-lived tasks, *e.g.* web apps that fulfil thousands of HTTP client requests by fetching static content or executing back-end commands.

**$C_M$ Moderate concurrency scenarios** engage in long-running, computationally-intensive tasks, *e.g.* Big Data stream processing frameworks.

Our benchmark workloads match the hardware capacity afforded by $P_E$ and $P_G$:

**High concurrency benchmarks** on $P_E$ set $n = 100\text{k}$ workers and $w = 100$ work requests per worker. These generate $\approx (n \times w \text{ requests} \times w \text{ responses}) = 20\text{M}$ message exchanges between the master and worker processes, totalling $\approx (20\text{M} \times \text{!} \text{ events} \times \text{?} \text{ events}) = 40\text{M}$ analysable trace events. Platform $P_G$ sets $n = 500\text{k}$ workers batched with $w = 100$ requests to produce $\approx 100\text{M}$ messages and $\approx 200\text{M}$ trace events. The high concurrency model $C_H$ is studied in sec. 5.4.

**Moderate concurrency benchmarks** on $P_G$ set $n = 5\text{k}$ workers and $w = 10\text{k}$ work requests per worker. These settings yield roughly the same number of trace events as on $P_G$ with concurrency scenario $C_H$. The moderate concurrency model $C_M$ is studied in sec. 5.5.

All experiments in secs. 5.4 and 5.5 use a total loading time of $t = 100$s. Each experiment consists of *ten* benchmarks that apply Steady, Pulse, and Burst workloads. We repeat every experiment *three* times to obtain *negligible variability* and ensure the accuracy of our results; see app. C.4 for a summary of these workloads and app. C.5 for the precautions we take.

The hardware, OS, and Erlang versions of platforms $P_E$ and $P_G$, combined with the workloads of concurrency scenarios $C_H$ and $C_M$ provide generality to our conclusions.

## 5.3 Instrumentation configuration

One challenge in conducting our experiments is the lack of RV monitoring tools targeting the EVM. To the best of our knowledge [65, Tables 3 and 4], detectEr [75, 16, 17, 15, 73, 40] is the only RV tool for Erlang that implements centralised outline instrumentation[3]. We are unaware of inline RV tools besides [39] and [3, 4]. Since the former tool is *unavailable*, we use the latter, more recent work[4]. In our experiments, we instrument the master *and each* worker process in the SuS models generated from sec. 5.2 to exert the highest possible load and capture *worst-case* scenarios. BenchCRV annotates work requests and responses with a unique sequence number to account for each message in benchmark runs. We leverage this numbering to write specialised monitor replicas that ascertain the *soundness* of trace event sequences reported to every RV monitor linked with the master and workers; see app. C.5 for details. Equally crucial, this runtime checking introduces a degree of *realistic* RV analysis slowdown that is *uniform* across all monitors in the inline, centralised, and RIARC monitoring set-ups. We empirically estimate this slowdown at $\approx 5\mu s$ per analysed event.

### 5.4   High concurrency benchmarks

We study runtime overhead in the high concurrency scenario $C_H$ with two aims. First, we show the effect overhead has on the SuS as it executes. Specifically, we consider how the memory consumption and scheduler utilisation impact the *latency* a client of the SuS experiences, *e.g.* end-user or application. We use the edge-case platform $P_E$ for these experiments; analogous results obtained on $P_G$ are detailed in app. C. Our second goal targets the general-case platform $P_G$ to assess the *scalability* of the instrumentation methods through their optimal use of the *additional* memory and scheduler capacity afforded by $P_G$.

The charts in secs. 5.4.1–5.4.3 plot performance metrics, *e.g.* memory consumption ($y$-axis) against the number of concurrent worker processes or the execution duration ($x$-axis). Since inline instrumentation prevents us from delineating the SuS and monitoring-induced runtime overhead, we follow the standard RV literature practice and include the *baseline* plots, *e.g.* [17, 75, 46, 39, 102, 117, 115]. Baseline plots show the *unmonitored* SuS to compare the relative overhead between each evaluated instrumentation method.

### 5.4.1   Instrumentation overhead

The first set of experiments isolates the instrumentation overhead induced on the SuS: this is the aggregated cost of tracing *and* reporting the traces soundly per def. 1 to RV monitors. Crucially, these experiments *omit monitors*, as we want to quantify the instrumentation overhead and understand its impact on the SuS. This enables us to focus on the differences between inlining—regarded as the most efficient instrumentation method [63, 62, 21]—and outlining. As far as we know [65, 74], outlining has *never* been used for decentralised RV in a *dynamic* setting such as ours. While we confirm that inline instrumentation uses less memory and scheduler capacity, RIARC dynamically scales and economises their use *without* adverse impact on the latency. In fact, the latency induced by RIARC is a mere 519ms higher than that of inline instrumentation at the peak stress-inducing loading point of 3.7k workers/s under Burst workloads. Our experiments indicate that centralised instrumentation manages resources poorly due to its inability to scale, increasing the chances of failure; see sec. 5.4.2.

Fig. 7 plots our results. Centralised instrumentation carries the largest overhead penalty. Regardless of the workload applied, it uses the most memory, $\approx 3.8$ GB, highlighting its ineptitude to scale. This stems from the backlog of trace event messages that accumulate in the mailbox of the central tracer and is a manifestation of two aspects. First, the central tracer does not consume events at the same rate worker processes produce them. Evidence of this *bottleneck* is visible as high scheduler utilisation in fig. 7 (bottom). This values settles at $\approx 36\%$ for the benchmarks with $\approx 40$k workers under the Steady workload and $\approx 60$k workers under Pulse and Burst workloads. Interpreting these $< 36\%$ scheduler usage values in isolation may suggest that centralised instrumentation has the potential to scale. However, its memory consumption plots in fig. 7 (middle) contradict this erroneous hypothesis.

---

[3] https://bitbucket.org/duncanatt/detecter-lite
[4] https://github.com/ScienceofComputerProgramming/SCICO-D-22-00294

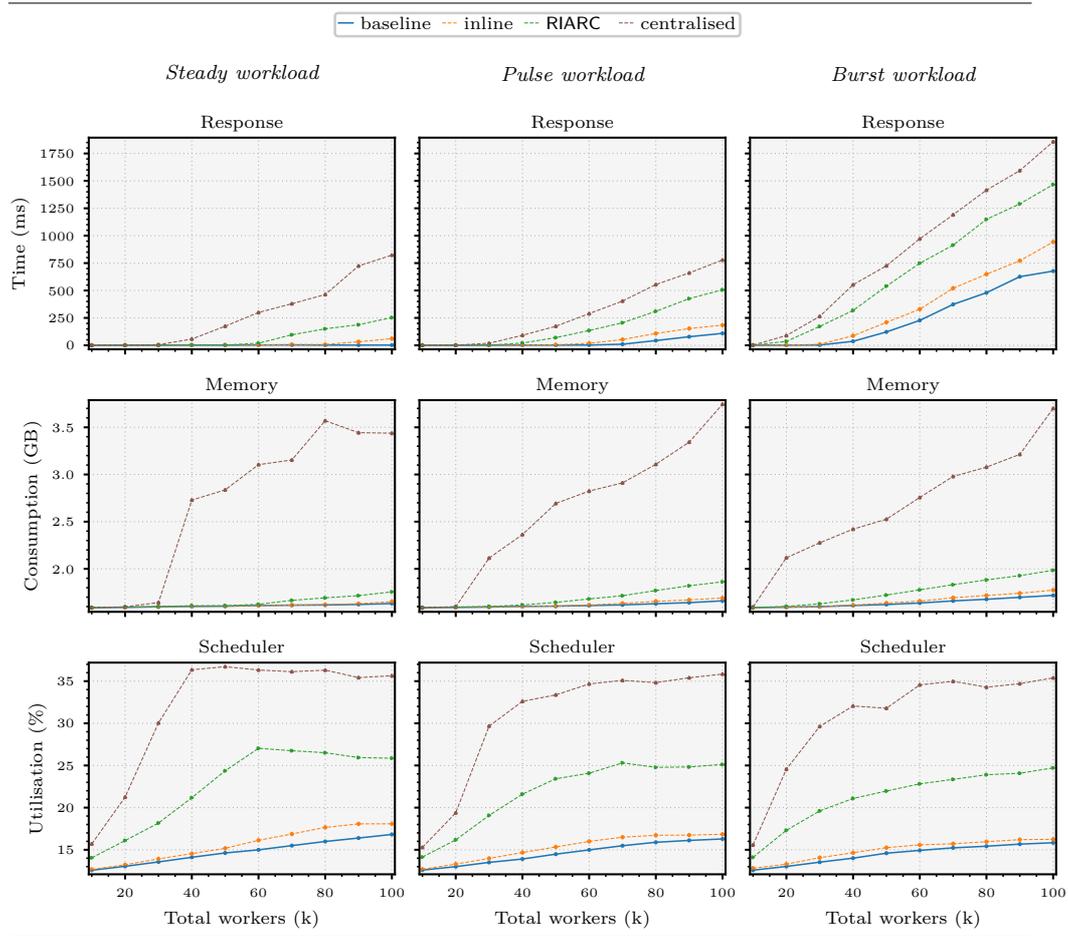

**Figure 7** Isolated instrumentation overhead (*high* workload, 100k workers)

By contrast, RIARC uses fewer resources to yield lower response times across the three workloads. The scheduler utilisation for RIARC slightly plateaus in the Steady ($\approx 60$k workers) and Pulse ($\approx 70$k workers) workload charts. This is not owed to scalability limitations of RIARC but to the intrinsic throttling instigated by the master process [127]. In fact, the plots for the baseline system and inline instrumentation in fig. 7 (middle) exhibit analogous signs of throttling. Even at a peak Burst workload of 3.7k workers/s, inline and RIARC instrumentation consume fairly similar amounts of memory, 1.7GB *vs.* 1.9GB, respectively.

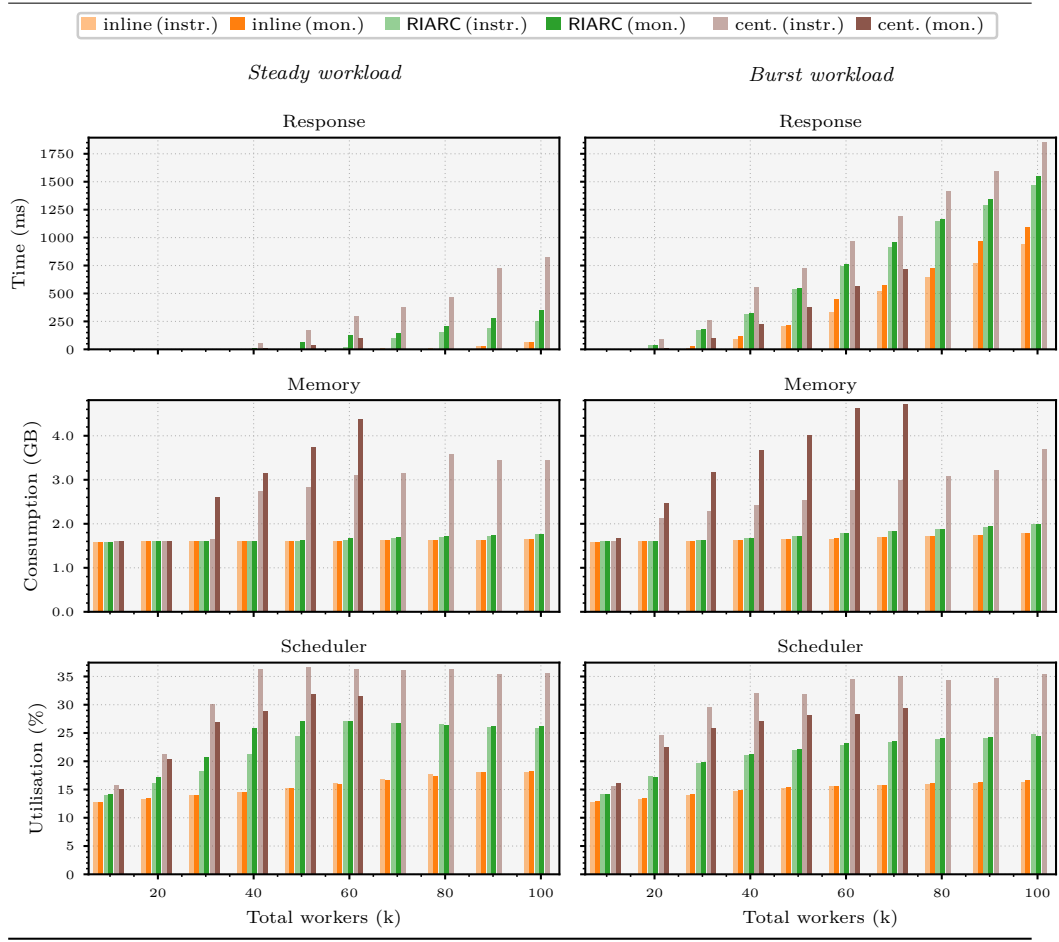

**Figure 8** Instrumentation and RV monitoring overhead gap (*high* workload, 100k workers)

### 5.4.2 Monitoring overhead

Our second set of experiments extends the results of sec. 5.4.1 and quantifies the cost of RV monitoring. The *runtime monitoring* overhead combines the instrumentation and slowdown due to the RV analysis, established at $\approx 5\mu s$ per event in sec. 5.3 for our experiments. Fig. 8 plots the instrumentation (*instr.*) overhead from sec. 5.4.1 next to the runtime monitoring overhead (*mon.*). It shows that the RV analysis slowdown aggravates centralised monitoring to the point of crashing. Inline and RIARC monitoring are minimally affected. Our results also reveal that the instrumentation incurs the *major* overhead portion, not the RV analysis. Sec. 5.6 comments on this finding in the context of existing RV tools.

Fig. 8 plots our results under the Steady and Burst workloads; fig. 14 in app. C.6.1 includes all three workloads. The charts for centralised monitoring exhibit a significant disparity between the instrumentation and runtime monitoring bar plots as the workload increases. This trend is consistent across both workloads in fig. 8. The lack of scalability of centralised monitoring in fig. 8 manifests as an increase in memory consumption but stabilised scheduler usage, as in fig. 7. Memory consumption and scheduler usage for centralised monitoring grow rapidly beyond $\approx 30$k and $\approx 20$k workers under the Steady and Burst workloads, respectively.

Bottlenecks led our experiments to crash (shown as missing bar plots in fig. 8). Crashes occur at $\approx$ 70k workers under the Steady and at $\approx$ 80k under Burst workload. By analysing the resulting dumps, we could attribute these crashes to memory exhaustion, which caused the EVM to fail. The dumps indicate severe memory pressure due to the vast backlog of trace event messages in the mailbox of the central tracer.

Inline and RIARC monitoring scale to accommodate the RV analysis slowdown. This is confirmed by cross-referencing the memory consumption and scheduler utilisation in fig. 8 for both monitoring methods. Each displays comparable overhead in their respective instrumentation and corresponding runtime monitoring bar plots. Fig. 8 (top) shows that inline and RIARC monitoring increase the latency, albeit for different reasons. The internal operation of RIARC enables us to deduce that its latency stems from message routing and dynamic tracer reconfiguration. Its scheduler utilisation plots support this observation. The latency due to inlining is a direct effect of RV analysis slowdown, provoked by the lock-step execution of monitors and the SuS. Other works, *e.g.* [46, 38], offer similar observations.

Dissecting our results uncovers further subtleties. The optimal scheduler utilisation of RIARC implies that its monitors are only active when triggered by trace events but remain idle otherwise. This inference is supported by the absence of sudden or continued memory growth for RIARC in fig. 8 (middle). The instrumentation and runtime monitoring latency bar plots for inline monitoring exhibit a growing pairwise gap that starts at $\approx$ 80k workers in fig. 8 (top right). The respective gap for RIARC at this mark is perceptibly lower. We credit this lower latency gap to outlining, which absorbs the slowdown effect of RV analyses. This leads us to conjecture that RIARC could accommodate monitors that perform richer RV analyses with minimal impact on the SuS. Our calculations from fig. 8 (top right) put the latency at 1093ms for inline monitoring *vs.* 1547ms for RIARC at a peak Burst workload of 3.7k workers/s: a 454ms difference, which is *lower* than the 519ms gap measured in sec. 5.4.1. Sec. 5.5 shows this gap is negligible in moderate concurrency scenarios.

### 5.4.3   Resource usage

We employ platform $P_G$ with high concurrency $C_H$ to confirm that our observations about inline and RIARC monitoring transfer to general cases. Secs. 5.4.1 and 5.4.2 deem centralised monitoring to be impractical. We, thus, omit it from the sequel; see app. C.6.3 for results.

Our experiments now use 16 scheduling threads, $n=500\text{k}$ workers, and $w=100$ requests per worker, producing $\approx$ 100M messages and $\approx$ 200M trace events. Fig. 13 in app. C.4 render these Steady, Pulse, and Burst workload models. Secs. 5.4.1 and 5.4.2 bound the memory and scheduler metrics to the period the SuS executes to portray the *actual overhead* impact on the system. We refocus that view to assess the monitoring overhead in *its entirety*—from the point of SuS launch until monitors complete their RV analysis. Doing so reveals how inline and RIARC monitoring optimise the use of added memory and processing capacity. Results show that inline and RIARC monitoring are elastic and dynamically adapt to changes in the applied workloads. App. C.6.3 reconfirms that centralised monitoring lacks this trait.

Fig. 9 gives a complete benchmark run under the Steady and Burst workloads. We relabel the *x*-axis with the benchmark duration and omit the response time plots since response time is inapplicable to these experiments (latency is an attribute of the SuS, not the monitors). In this run, the Steady workload generates a sustained load of $\approx$ 5k workers/s whereas Burst peaks at $\approx$ 17.8k workers/s under maximum load at $\approx$ 5s; see fig. 13 in app. C.4.

Fig. 9 (top) illustrates the memory consumption patterns for inline and RIARC monitoring, which exhibit *elasticity*. This elastic behaviour occurs at different points in the plots. Inline monitoring peaks at $\approx$ 3.7GB at $\approx$ 72s and RIARC at $\approx$ 5.7GB at $\approx$ 100s under the Burst workload. The memory consumption for both methods stabilises at around $\approx 36$s under the Steady workload, with $\approx 2.3$GB for inline and $\approx 2.7$GB for RIARC monitoring. Elasticity in these methods is due to different reasons: it is intrinsic to inline monitoring (see sec. 1), whereas the RIARC spawns and garbage collects monitors on demand (secs. 3.1 and 3.6). Fig. 16 in app. C.6.3 certifies these observations under the Pulse workload. Centralised monitoring is *insensitive* to the workload applied, as figs. 17 and 18 in app. C.6.3 reconfirm.

The effect of dynamic message routing and tracer reconfiguration that RIARC performs is evident in the scheduler utilisation plots of fig. 9. Under the Steady and Burst workloads, scheduler utilisation oscillates continually due to the sustained influx of trace events. Oscillations corroborate our observation in sec. 5.4.2 about RIARC, namely, that monitors are activated by trace events but remain idle otherwise. Active monitor periods manifest as peaks in fig. 9. Idle periods, where monitors are placed in the EVM waiting queues, are reflected as regions with low and stable scheduler utilisation. These oscillations showcase the message-driven aspect of RIARC, which analyses events asynchronously. Inlining exhibits minimal scheduler utilisation oscillations due to its lock-step execution with the SuS.

## 5.5 Moderate concurrency benchmarks

Our last experiment studies moderate concurrency scenarios $C_M$. The general-case platform $P_G$ sets $n = 5$k workers and $w = 10$k requests per worker, and uses 16 EVM schedulers. We show that under these loads, RIARC induces overhead on par with inline monitoring.

Moderate concurrency alters the execution of the master-worker model, compared to our benchmarks of secs. 5.4.1–5.4.3. In this set-up, the master creates most of its worker processes at the initial stage of benchmark runs and spends the remaining time allocating work requests. This change grows the request throughput markedly, *e.g.* see tbl. 5 in app. C.4. One consequence is that centralised monitoring consistently crashes under the rapid accumulation of messages in its mailbox. We, thus, limit our study to inline and RIARC monitoring.

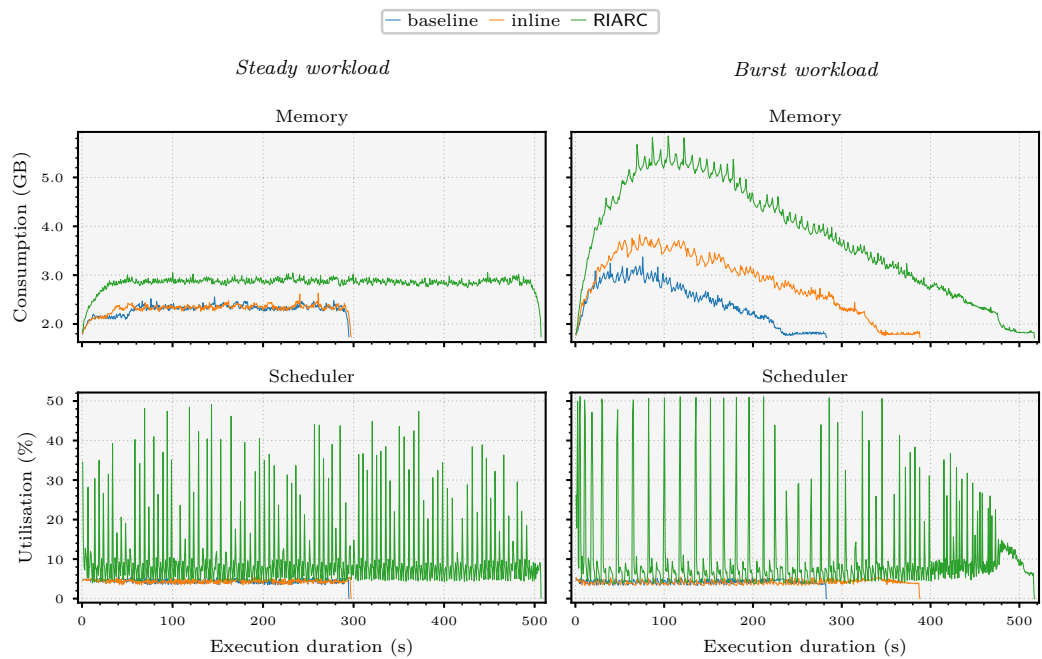

**Figure 9** Inline and RIARC monitoring resource usage (*high* workload, 500k workers)

Tbl. 3 compares the results taken on platform $P_G$ from sec. 5.4.3 with 500k workers (high concurrency, $C_H$) against the ones on $P_G$ with 5k workers (moderate concurrency, $C_M$). The figures shown estimate the percentage overhead w.r.t. the baseline systems $C_H$ and $C_M$ at this *maximum* load. Our ensuing discussion is limited to the overhead under the Steady and Burst workloads since each respectively captures the SuS operation in *typical* and *severe* load conditions. Readers are referred to fig. 20 in app. C.6.4 for the overhead comparison given in absolute metric values for the entirety of benchmark runs.

Tbl. 3 indicates that the memory consumption overhead due to inline monitoring is not affected under the Steady workload, which remains at 1% in both the high and moderate concurrency scenarios $C_H$ and $C_M$. However, it decreases from 16% in $C_H$ to 1% in $C_M$. We observe the opposite effect on the scheduler utilisation overhead for inline monitoring. For the moderate concurrency case $C_M$, the scheduler overhead under the Steady and Burst workloads increases to 3% and 4% respectively.

Tbl. 3 also shows that under the Steady workload, RIARC induces a 23% memory overhead in concurrency scenario $C_H$ *vs.* 8% in concurrency scenario $C_M$, a decrease of 15%. Under the Burst workload, this overhead is reduced by 46%, from 56% in $C_H$ to 10% in $C_M$. The scheduler utilisation overhead for RIARC from $C_H$ to $C_M$ also registers drops of ≈71% under both Steady and Burst workloads. We attribute these overhead improvements to the lower number of worker processes the master creates in the moderate concurrency set-up, $C_M$. The long-running worker processes induce stability in the SuS. RIARC adapts to this change favourably by performing fewer trace event routing and tracer reconfigurations. The ramification of this adaptability is perceivable in the latency overhead discussed next.

RIARC inflates the latency overhead from 95% in $C_H$ to 194% in $C_M$ under the Steady workload (+99%), and from 97% in $C_H$ to 190% in $C_M$ under the Burst workload (+93%). However, RIARC induces *less latency* overhead than inline monitoring. Tbl. 3 reveals that the latency overhead for inline monitoring grows from 4% in the high concurrency set-up $C_H$ to 246% in the moderate concurrency set-up $C_M$ under the Steady workload (+242%). It also grows under the Burst workload, from 55% in $C_H$ to 193% in $C_M$ (+138%). In fact, our calculations confirm that the *absolute* response time for inline monitoring is slightly worse than that of RIARC in $C_M$: 116ms *vs.* 98ms under the Steady, and 182ms *vs.* 179ms under the Burst workloads respectively. This latency degradation for inline monitoring stems from the ≈5µs slowdown induced by the RV analysis, which results in frequent 'pausing' of worker processes. Monitors comprising richer analyses produce longer pauses in worker processes, which can degrade the response time further [46, 38, 72].

| Concurrency | Workload | Response time % | | Memory consumption % | | Scheduler utilisation % | |
|---|---|---|---|---|---|---|---|
| | | Inline | RIARC | Inline | RIARC | Inline | RIARC |
| $C_H$ (500k) | Steady | 4 | 95 | 1 | 23 | 0 | 123 |
| | Burst | 55 | 97 | 16 | 56 | 0 | 123 |
| $C_M$ (5k) | Steady | 246 | 194 | 1 | 8 | 3 | 52 |
| | Burst | 193 | 190 | 1 | 10 | 4 | 50 |

**Table 3** Percentage overhead on $C_H$ (500k) and $C_M$ (5k) w.r.t. baseline at *maximum* workload

## 5.6  Discussion

The RIARC scheduler utilisation in tbl. 3 is higher than the reported values for inline monitoring. This should not be construed as an inefficiency. From a reactive systems perspective, growth in the scheduler utilisation indicates *scalability*, as the low memory consumption in tbl. 3 affirms. RIARC benefits from the ample schedulers to improve the overall system response time *without* overtaxing the system. Indeed, fig. 20 in app. C.6.4 demonstrates that the mean absolute scheduler utilisation in the benchmarks of sec. 5.5 is just $\approx 10\%$ under both the Steady and Burst workloads. Tbl. 3 shows that the reduction in latency makes RIARC comparable to inline monitoring in moderate concurrency scenarios.

Sec. 1 names *responsiveness* as a key reactive systems attribute [97]. RIARC prioritises responsiveness by isolating its monitors into asynchronous concurrent units. This design naturally exploits the available processing capacity of the host platform by maximising monitor *parallelism* when possible. Inline monitoring reaps fewer benefits in identical settings because its lock-step execution with the SuS robs it of potential parallelism gains.

Secs. 5.4.1 – 5.4.3 attest to the impracticality of centralised monitoring for reactive systems. Bottlenecks hinder its ability to scale, compelling it to consume inordinate amounts of memory, which can lead to failure, as sec. 5.4.2 shows. Despite these shortcomings, many RV tools in this setting use centralised monitoring, *e.g.* [50, 16, 133, 66, 84, 113, 75, 38, 41, 39, 2, 106].

## 6  Conclusion

Reactive software calls for instrumentation methods that uphold the responsive, resilient, message-driven, and elastic attributes of systems. This is attainable *only if* the instrumentation exhibits these qualities. Runtime verification imposes another demand on the instrumentation: that the trace event sequences it reports to monitors are *sound*, *i.e.,* traces do not omit events and preserve the ordering with which events occur locally at processes.

This paper presents RIARC, a novel decentralised instrumentation algorithm for outline monitors meeting these two demands. RIARC uses outline monitors to decouple the runtime analysis from system components, which minimises latency and promotes *responsiveness*. Outline monitors can fail independently of the system and each other to improve *resiliency*. RIARC gathers events non-invasively via a tracing infrastructure, making it *message-driven* and suited to cases where inlining is inapplicable. The algorithm is *elastic*: it reacts to specific events in the trace to instrument and garbage collect monitors on demand.

Our asynchronous setting complicates the instrumentation due to potential trace event loss or reordering. RIARC overcomes these challenges using a next-hop IP routing approach to rearrange and report events soundly to monitors. We validate RIARC by subjecting its corresponding Erlang implementation to rigorous systematic testing, confirming its correctness. This implementation is evaluated via extensive empirical experiments. These subject the implementation to large realistic workloads to ascertain its reactiveness. Our experiments show that RIARC optimises its memory and scheduler usage to maintain latency feasible for soft real-time applications. We also compare RIARC to inline and centralised monitoring, revealing that it induces *comparable* latency to inlining under moderate concurrency.

**Related work**   Works on inlining besides the ones cited in sec. 1, *e.g.* [81, 25, 50, 49, 53, 52], do not separate the instrumentation and runtime analysis. This is common in monolithic settings, where the instrumentation is often assumed to induce minimal runtime overhead. As a result, many inline approaches focus on the efficiency of the analysis but neglect the instrumentation cost (*e.g.* [64] attributes overhead solely to the analysis). Sec. 5.4.1 shows this is not the case. This line of reasoning for monolithic systems is often ported to concurrent settings. For instance, [110, 133, 29, 46, 132, 67, 19] propose efficient runtime monitoring algorithms but do not account for, nor quantify, the overhead due to gathering trace events. Tools, such as [41, 38, 17, 35, 75, 142], that quantify the runtime overhead coalesce the instrumentation and runtime analysis costs, making it difficult to gauge whether inefficiencies arise from one or the other. We are unaware of empirical studies such as ours that distinguish between the instrumentation and runtime analysis overhead.

Sec. 5.6 remarks that centralised monitoring is used for concurrent runtime verification despite its evident limitations. One plausible reason for this is that the empirical scrutiny of such tools lacks proper benchmarking (*e.g.* [50, 16, 133, 66, 84]) or uses insufficient workloads that fail to expose the issues of centralised set-ups (*e.g.* [113, 75, 38, 41, 39, 2, 106]). Gathering inadequate metrics can also bias the interpretation of empirical data; see sec. 5.4.1. Works, such as [39, 17, 35, 131], consider the memory consumption and latency metrics. Our evaluation of inline, centralised, and RIARC monitoring uses (i) *combinations* of hardware and software, with (ii) two concurrency models that test *edge-case* and *general-case* scenarios, under (iii) *high* workloads that go beyond the state of the art, applying (iv) *realistic* workload profiles, interpreted against (v) *relevant* performance metrics that give a multi-faceted view of runtime overhead. To the best of our knowledge, this is generally not done in other studies, *e.g.* [117, 116, 47, 46, 124, 30, 109, 39, 41, 17, 50, 51, 53, 75, 60, 61, 27, 113, 100, 35].

Outline instrumentation decouples the execution of the SuS and monitor components in space (*i.e.,* isolated threads) and time (*i.e.,* asynchronous messaging). The tracing infrastructure outline instrumentation uses mirrors the publish-subscribe (Pub/Sub) pattern [138]. In this set-up, consumers subscribe to a *broker* that advertises events. Centralised instrumentation follows a Pub/Sub approach: the SuS produces trace events and deposits them into *one* global trace buffer that tracers receive from (see fig. 1b). Despite similarities, *e.g.* tracers register and deregister with the tracing infrastructure at runtime, RIARC differs from conventional Pub/Sub messaging in three fundamental aspects. Chiefly, Pub/Sub publishers are unaware of the subscribers interested in receiving messages because this bookkeeping task is appointed to the broker. By contrast, next-hop routing relies on the *explicit* address of recipients to forward messages. Furthermore, in Pub/Sub messaging, subscribers do not communicate with publishers, whereas RIARC tracers exchange *direct* detach requests between one another to reorganise the choreography (refer to sec. 3.4). Lastly, Pub/Sub brokers are typically predefined and remain fixed, while trace partitioning *reconfigures* the tracing topology, creating and destroying brokers in reaction to dynamic changes in SuS.

One assumption we make about process tracing is $A_4$, *i.e.,* tracing gathers the spawn events of parent processes before all the events of child processes. While $A_4$ induces a partial order over trace events, it is *weaker* than happened-before causality [98], as the events gathered from sets of child SuS processes need not be causally ordered. Demanding the latter condition would entail additional computation on the part of the tracing infrastructure and could increase runtime overhead. Maintaining minimal overhead is critical to our instrumentation because it preserves the responsiveness attribute of reactive systems. Tracing assumption $A_4$ and the RIARC logic detailed in sec. 3 guarantee trace soundness (def. 1), which suffices for RV monitoring. Since our work targets soft real-time systems [97, 95] scoped in a reliable messaging setting (see sec. 1), we do not tackle the problem of ensuring time-bounded causally-ordered message delivery [18] nor implement exactly-once delivery semantics [86]. We will address these challenges in future extensions of this work.

**L. Aceto, D. P. Attard, A. Francalanza, and A. Ingólfsdóttir** 33**116** Rumyana Neykova and Nobuko Yoshida. Let it Recover: Multiparty Protocol-Induced Recovery. In *CC*, pages 98–108, 2017.
**117** Rumyana Neykova and Nobuko Yoshida. Multiparty Session Actors. *LMCS*, 13, 2017.
**118** Nicolas Niclausse. Tsung, 2017. URL: http://tsung.erlang-projects.org.
**119** Scott Oaks. *Java Performance: In-Depth Advice for Tuning and Programming Java 8, 11, and Beyond.* CRC, 2020.
**120** Martin Odersky, Lex Spoon, Bill Venners, and Frank Sommers. *Programming in Scala.* Artima Inc., 2021.
**121** Athanansios Papoulis. *Probability, Random Variables, and Stochastic Processes.* McGraw Hill, 1991.
**122** Aleksandar Prokopec, Andrea Rosà, David Leopoldseder, Gilles Duboscq, Petr Tuma, Martin Studener, Lubomír Bulej, Yudi Zheng, Alex Villazón, Doug Simon, Thomas Würthinger, and Walter Binder. Renaissance: Benchmarking Suite for Parallel Applications on the JVM. In *PLDI*, pages 31–47, 2019.
**123** Kevin Quick. Thespian, 2020. URL: https://thespianpy.com/doc.
**124** Giles Reger, Helena Cuenca Cruz, and David E. Rydeheard. MarQ: Monitoring at Runtime with QEA. In *TACAS*, volume 9035 of *LNCS*, pages 596–610, 2015.
**125** Giles Reger, Sylvain Hallé, and Yliès Falcone. Third International Competition on Runtime Verification - CRV 2016. In *RV*, volume 10012 of *LNCS*, pages 21–37, 2016.
**126** Giles Reger and David E. Rydeheard. From First-Order Temporal Logic to Parametric Trace Slicing. In *RV*, volume 9333 of *LNCS*, pages 216–232, 2015.
**127** Sartaj Sahni and George L. Vairaktarakis. The Master-Slave Paradigm in Parallel Computer and Industrial Settings. *J. Glob. Optim.*, 9:357–377, 1996.
**128** Raja R. Sambasivan, Ilari Shafer, Jonathan Mace, Benjamin H. Sigelman, Rodrigo Fonseca, and Gregory R. Ganger. Principled Workflow-Centric Tracing of Distributed Systems. In *SoCC*, pages 401–414. ACM, 2016.
**129** Torben Scheffel and Malte Schmitz. Three-Valued Asynchronous Distributed Runtime Verification. In *MEMOCODE*, pages 52–61, 2014.
**130** Fred B. Schneider. Enforceable Security Policies. *ACM Trans. Inf. Syst. Secur.*, 3:30–50, 2000.
**131** Joshua Schneider, David A. Basin, Frederik Brix, Srdan Krstic, and Dmitriy Traytel. Scalable Online First-Order Monitoring. *Int. J. Softw. Tools Technol. Transf.*, 23:185–208, 2021.
**132** Koushik Sen, Grigore Rosu, and Gul Agha. Runtime Safety Analysis of Multithreaded Programs. In *ESEC / SIGSOFT FSE*, pages 337–346, 2003.
**133** Koushik Sen, Grigore Rosu, and Gul Agha. Online Efficient Predictive Safety Analysis of Multithreaded Programs. *Int. J. Softw. Tools Technol. Transf.*, 8:248–260, 2006.
**134** Koushik Sen, Abhay Vardhan, Gul Agha, and Grigore Rosu. Efficient Decentralized Monitoring of Safety in Distributed Systems. In *ICSE*, pages 418–427, 2004.
**135** Andreas Sewe, Mira Mezini, Aibek Sarimbekov, and Walter Binder. DaCapo con Scala: design and analysis of a Scala benchmark suite for the JVM. In *OOPSLA*, pages 657–676, 2011.
**136** SPEC. SPECjvm2008, 2008. URL: https://www.spec.org/jvm2008.
**137** Eric Stenman. *The Erlang Runtime System.* 2023.
**138** Sasu Tarkoma. *Overlay Networks: Toward Information Networking.* Auerbach, 2010.
**139** The Pony Team. Ponylang, 2021. URL: https://tutorial.ponylang.io.
**140** Ulf T. Wiger, Gösta Ask, and Kent Boortz. World-Class Product Certification using Erlang. *ACM SIGPLAN Notices*, 37(12):25–34, 2002.
**141** Jiali Yao, Zhigeng Pan, and Hongxin Zhang. A Distributed Render Farm System for Animation Production. In *ICEC*, volume 5709 of *LNCS*, pages 264–269, 2009.
**142** Teng Zhang, Greg Eakman, Insup Lee, and Oleg Sokolsky. Overhead-Aware Deployment of Runtime Monitors. In *RV*, volume 11757 of *LNCS*, pages 375–381, 2019.

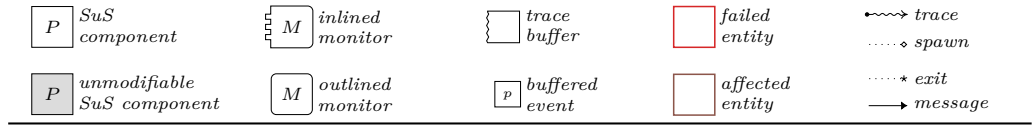

**Figure 10** Legend and notation for figures

## A   Appendix A: Auxiliary instrumentation logic

The operations $\textsc{Dispatch}(m,\imath_{\textsc{t}})$ and $\textsc{Forwd}(r,\imath_{\textsc{t}})$ given in alg. 4 enable tracers to perform next-hop routing, as described in sec. 3. $\textsc{Dispatch}$ embeds an evt or dtc acknowledgement message $m$ into a rtd packet, which is sent to the next-hop tracer with PID $\imath_{\textsc{t}}$. In the packet, $\textsc{Dispatch}$ also inserts the PID of the invoker tracer, obtained via the function self(). This is the PID of the *dispatch tracer*, and is used when a *forwarded* ⟡ event results in the instrumentation of a new SuS process (line 20 in alg. 3). Upon instrumenting the SuS PID carried by ⟡, the tracer issues a dtc request to that dispatch tracer PID. The function $\textsc{Detach}(\imath_{\textsc{s}},\imath_{\textsc{t}})$ encapsulates the detachment logic. It signals the dispatch tracer with PID $\imath_{\textsc{t}}$ that the SuS PID $\imath_{\textsc{s}}$ is being traced by the *current* tracer with PID $\jmath_{\textsc{t}} = \mathsf{self}()$; see line 13 in alg. 2 and line 13 in alg. 4. Before sending the dtc request, $\textsc{Detach}$ uses $\textsc{Preempt}$ so that the current tracer $\jmath_{\textsc{t}}$ takes over the tracing of SuS PID $\imath_{\textsc{s}}$. $\textsc{Forward}(r,\imath_{\textsc{t}})$ passes on the specified rtd packet $r$ to the next-hop, $\imath_{\textsc{t}}$. $\textsc{TryGC}$ determines whether a tracer can be safely terminated by confirming that the traced-processes and routing maps for a tracer are both empty.

Alg. 4 also includes the function $\textsc{Tracer}$ used by alg. 2 to spawn the core logic of algs. 1 and 3 to execute in a separate tracer process. $\textsc{Tracer}$ accepts four parameters:

1. $\sigma$, the state of the parent tracer,
2. $\varsigma_{\textsc{m}}$, the RV monitor signature utilised by the function $\textsc{AnalyseEvt}$ in algs. 1 and 3 to analyse trace events incrementally,
3. $\imath_{\textsc{s}}$, the PID of the SuS process to instrument, and
4. $\imath_{\textsc{t}}$, the PID of the dispatch tracer (from the rtd packet) to which the dtc request is issued.

The process tracing functions $\textsc{Trace}$, $\textsc{Clear}$ and $\textsc{Preempt}$ described in sec. 3 are listed in alg. 5. $\textsc{Trace}$ and $\textsc{Clear}$ abstract the inner workings of the EVM tracing exposed via the

**Algorithm 4** Operations used by the direct (○) and priority (●) tracer loops

**Expect:** $m = \langle \mathsf{evt}, \ell, \imath_{\textsc{s}}, \jmath_{\textsc{s}}, \varsigma_{\textsc{s}} \rangle \vee m = \langle \mathsf{dtc}, \imath_{\textsc{t}}, \imath_{\textsc{s}} \rangle$

1  **def** $\textsc{Dispatch}(m, \imath_{\textsc{t}})$
2    $\imath_{\textsc{t}} \,!\, \langle \mathsf{rtd}, \mathsf{self}(), m \rangle$

3  **def** $\textsc{Detach}(\imath_{\textsc{s}}, \imath_{\textsc{t}})$
4    $\jmath_{\textsc{t}} \leftarrow \mathsf{self}()$
5    $\textsc{Preempt}(\imath_{\textsc{s}}, \jmath_{\textsc{t}})$   *# This tracer takes over*
6    $\imath_{\textsc{t}} \,!\, \langle \mathsf{dtc}, \jmath_{\textsc{t}}, \imath_{\textsc{s}} \rangle$

**Expect:** $r = \langle \mathsf{rtd}, \imath_{\textsc{t}}, m \rangle$

7  **def** $\textsc{Forwd}(r, \imath_{\textsc{t}})$
8    $\imath_{\textsc{t}} \,!\, r$

9  **def** $\textsc{TryGC}(\sigma)$
10   **if** $(\sigma.\Gamma = \emptyset \wedge \sigma.\Pi = \emptyset)$ Terminate tracer

11 **def** $\textsc{Tracer}(\sigma, \varsigma_{\textsc{m}}, \imath_{\textsc{s}}, \imath_{\textsc{t}})$
   *# New tracer state $\sigma'$ initialised with:*
   *# 1. empty routing map, $\emptyset$*
   *# 2. copy of instrumentation map, $\sigma.\Lambda$*
   *# 3. traced-process map with **first** process*
   *#    to trace, $\imath_{\textsc{s}}$*
12   $\sigma' \leftarrow \langle \Pi \leftarrow \emptyset, \sigma.\Lambda, \Gamma \leftarrow \{\langle \imath_{\textsc{s}}, \bullet \rangle\} \rangle$
   *# Issue dtc request for SuS PID $\imath_{\textsc{s}}$*
   *# to dispatch tracer $\imath_{\textsc{t}}$*
13   $\textsc{Detach}(\imath_{\textsc{s}}, \imath_{\textsc{t}})$

   *# Start with empty trace buffer $\kappa$ and in*
   *# ● mode to prioritise forwarded messages*
14   $\textsc{Loop}_{\bullet}(\sigma', \varsigma_{\textsc{m}})$

▎ **Algorithm 5** Abstraction of the operations offered by process tracing

| | |
|---|---|
| 1 **def** TRACE($\imath_S$,$\imath_T$) | 9 **def** CLEAR($\imath_S$,$\imath_T$) |
| 2  **if** ($\imath_S$ is **not** traced) | 10  **if** ($\imath_S$ is traced) |
| 3   Set tracer for SuS PID $\imath_S$ to $\imath_T$ | 11   Clear tracer $\imath_T$ from SuS PID $\imath_S$ |
|    *# Child processes of $\imath_S$, their children, etc.* |    *# Child processes of $\imath_S$, their children, etc.* |
|    *# inherit $\imath_T$, tracing assumption $A_2$* |    *# still traced by $\imath_T$, tracing assumption $A_2$* |
| 4  **while** tracer of $\imath_S$ is set **do** | 12  **repeat** |
|    *# Read details of next trace event of $\imath_S$* | 13   no-op |
| 5   $\ell,\imath_S,\jmath_S,\varsigma_S \leftarrow$ trace event exhibited by $\imath_S$ | 14  **until** events of $\imath_S$ **are** delivered to $\imath_T$ |
|    *# Encode details as message, see sec. 2.2* | 15 **def** PREEMPT($\imath_S$,$\imath_T$) |
| 6   $e = \langle \mathsf{evt},\ell,\imath_S,\jmath_S,\varsigma_S \rangle$ | 16  $\imath'_T \leftarrow$ current tracer of SuS PID $\imath_S$ |
| 7   $\imath_T\,!\,e$ *# Send event to trace buffer of $\imath_T$* | 17  CLEAR($\imath_S$,$\imath'_T$) *# Tracer $\imath'_T$ stops tracing $\imath_S$* |
| 8  **end while** | 18  TRACE($\imath_S$,$\imath_T$) *# Tracer $\imath_T$ starts tracing $\imath_S$* |

Erlang built-in primitive `trace`, and the underlying operation of our offline tracing engine described in sec. 4.1 and app. B.

The function START in alg. 6 launches the SuS and root tracer in tandem. START accepts the main SuS function signature $\varsigma_S$ together with the instrumentation map, $\Lambda$. *Copies* of this map (see line 12 in alg. 4) are propagated between tracers, enabling them to determine whether a spawned SuS process requires instrumentation through a separate tracer. To safeguard against the initial loss of trace events, the SuS is launched in a *paused* state (line 2). This permits the root tracer to start tracing the root system process that runs $\varsigma_S$. ROOT resumes the system (line 6), and begins its trace inspection in *direct* mode, as line 8 shows.

▎ **Algorithm 6** Launching root SuS and tracer processes

| | |
|---|---|
| 1 **def** START($\varsigma_S$,$\Lambda$) | 4 **def** ROOT($\imath_S$,$\Lambda$) |
|    *# Pausing allows root tracer to be set* | 5  TRACE($\imath_S$,self()) |
|    *# up; no initial message loss* | 6  Resume root SuS process with PID $\imath_S$ |
| 2  $\imath_S \leftarrow \mathsf{spwn}(\varsigma_S)$ in paused mode | 7  $\sigma \leftarrow \langle \Pi \leftarrow \emptyset, \Lambda, \Gamma \leftarrow \{\langle \imath_S,\circ \rangle\} \rangle$ |
| 3  $\imath_T \leftarrow \mathsf{spwn}(\mathrm{ROOT}(\imath_S,\Lambda))$ | 8  LOOP$_\circ(\sigma,\bot)$ |

## B  Appendix B: Offline tracing and algorithm invariants

RIARC can be extended with the event reordering scheme described when the underlying tracing infrastructure does not guarantee tracing assumption $A_4$. This can be done in Erlang by peeking at the mailbox using the built-in primitive `process_info`. In principle, this is inefficient if the mailbox contains many messages [42]. We, however, remark that in practice, such inefficiency arises only in the extreme case where ◇ events are deposited into a tracer mailbox in exactly the reverse order in which descendant processes are spawned. Alternatively, one can use an auxiliary trace buffer (*e.g.* a list) that is populated by dequeuing the tracer mailbox first. Both amendments can be made on lines 3 of algs. 1 and 3.

### B.1  Offline tracing

Ex. 7 sketches below how our offline tracing engine operates. Internally, it uses tracer buffers and sets of processes to rearrange process ◇ events for descendant SuS processes. The tracing engine rearranges ◇ events using the PID information they carry. In doing so, it recovers the happens-before causality between each ◇ event. Concurrent ◇ events for sibling processes, such as when process $P$ spawns $Q$ and $R$, are not reordered.

▶ **Example 7** (Reordering spawn events). Suppose the tracer $T_P$ with PID $p_\text{T}$ registers to trace the SuS process $P$ with PID $p_\text{S}$. $P$ spawns process $Q$, which, in turn, spawns $R$, as in fig. 5a. $T_P$ invokes $\text{TRACE}(p_\text{S}, p_\text{T})$, which registers its PID $p_\text{T}$ with the tracing engine. The tracing engine assigns the empty trace buffer $B$ and set $S = \{p_\text{S}\}$ to $p_\text{T}$.

**Scan 1.** When the event $e_1 = \langle \text{evt}, ?, q_\text{S} \rangle$ is read into $B$, the engine does not deliver it to $p_\text{T}$. The occurs because none of the SuS PID values in $S$ match the value of the originator PID in the $?_Q$ event, *i.e.*, $e_1.\imath_\text{S} = q_\text{S} \notin \{p_\text{S}\}$.
**Scan 2.** Event $e_2 = \langle \text{evt}, \diamond, q_\text{S}, r_\text{S}, f_{s_R} \rangle$ is read next into the buffer. A scan is performed but no action is taken, as $e_2.\imath_\text{S} = q_\text{S} \notin \{p_\text{S}\}$. $B$ now contains '$?_Q.\diamond_Q$'.
**Scan 3.** Events $e_3 = \langle \text{evt}, \diamond, p_\text{S}, q_\text{S}, f_{s_Q} \rangle$ and $e_4 = \langle \text{evt}, !, p_\text{S}, q_\text{S} \rangle$ are appended to $B$. The engine scans $B$ and dequeues $\langle \text{evt}, \diamond, p_\text{S}, q_\text{S}, f_{s_Q} \rangle$ since the value of the originator PID $e_3.\imath_\text{S} = p_\text{S}$ is contained in $\{p_\text{S}\}$. This triggers the event $\diamond_P$ to be delivered to $T_P$. Additionally, the engine sets $S = \{p_\text{S}, q_\text{S}\}$ per the inheritance tracing assumption $A_2$ of sec. 2.
**Scan 4.** Updating $S$ triggers another buffer scan to check whether any events require dequeuing. The event $\langle \text{evt}, ?, q_\text{S} \rangle$ is dequeued and delivered to $T_P$, since now, $e_1.\imath_\text{S} = q_\text{S} \in \{p_\text{S}, q_\text{S}\}$. Similarly, $\langle \text{evt}, \diamond, q_\text{S}, r_\text{S}, f_{s_R} \rangle$ is dequeued and delivered to $T_P$. $S$ is updated to $\{p_\text{S}, q_\text{S}, r_\text{S}\}$. The engine continues scanning the buffer and dequeues $\langle \text{evt}, !, p_\text{S}, q_\text{S} \rangle$, which it delivers to $T_P$.

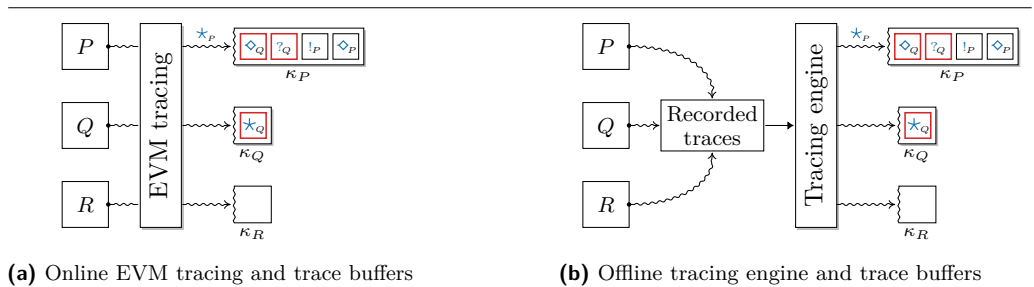

**(a)** Online EVM tracing and trace buffers    **(b)** Offline tracing engine and trace buffers

▮ **Figure 11** Online tracing via the EVM and offline tracing based on replayed trace files

**Scan 5.** Since $B$ is empty, the update in $S$ does not trigger another buffer scan. The engine pauses until new events are read into the buffer.

The input trace in the buffer '$?_Q.\diamond_Q.\diamond_P.!_P$' has been delivered to $T_P$ as '$\diamond_P.?_Q.\diamond_Q.!_P$', matching the one shown in fig. 5a. ◂

▶ **Example 8** (Other interleaved executions). Other executions are possible. The input buffer '$?_Q.\diamond_P.\diamond_Q.!_P$' results in the same trace '$\diamond_P.?_Q.\diamond_Q.!_P$' of fig. 5a reaching $T_P$. ◂

We underscore that the *input* traces '$?_Q.\diamond_Q.\diamond_P.!_P$' and $?_Q.\diamond_P.\diamond_Q.!_P$ from exs. 7 and 8 observe *trace consistency* of def. 1 w.r.t. $P$ and $Q$. For instance, the input trace '$\diamond_Q.?_Q.\diamond_P.!_P$' is inconsistent w.r.t. $Q$. Ex. 9 shows that our tracing engine preserves *trace identity*, *i.e.*, a consistent trace with the correct causal ordering between $\diamond$ events in descendant SuS processes is not modified.

▶ **Example 9** (Trace identity). For the same tracer set-up of ex. 7, *i.e.*, $T_P$ initially tracing $P$, the buffer '$e_1 = \langle \mathsf{evt},\diamond,p_\mathsf{s},q_\mathsf{s},f_{\mathsf{s}_Q}\rangle.e_2 = \langle \mathsf{evt},?,q_\mathsf{s}\rangle.e_3 = \langle \mathsf{evt},!,p_\mathsf{s},q_\mathsf{s}\rangle.e_4 = \langle \mathsf{evt},\diamond,q_\mathsf{s},r_\mathsf{s},f_{\mathsf{s}_R}\rangle$', and $T = \{p_\mathsf{s}\}$, our trace engine performs the following scans:

**Scan 1.** Event $e_1 = \langle \mathsf{evt},\diamond,p_\mathsf{s},q_\mathsf{s},f_{\mathsf{s}_Q}\rangle$ is read and delivered to $T_P$ since $e_1.\imath_\mathsf{s} = p_\mathsf{s} \in \{p_\mathsf{s}\}$. $T$ is updated to $\{p_\mathsf{s},q_\mathsf{s}\}$, by tracing assumption $A_2$.
**Scan 2.** The update in $T$ triggers the next scan. Event $e_2 = \langle \mathsf{evt},?,q_\mathsf{s}\rangle$ is delivered to $T_P$, as $e_2.\imath_\mathsf{s} = q_\mathsf{s} \in \{p_\mathsf{s},q_\mathsf{s}\}$. The events $\langle \mathsf{evt},!,p_\mathsf{s},q_\mathsf{s}\rangle$ and $\langle \mathsf{evt},\diamond,q_\mathsf{s},r_\mathsf{s},f_{\mathsf{s}_R}\rangle$ follow, and $T$ is updated to $\{p_\mathsf{s},q_\mathsf{s},r_\mathsf{s}\}$.
**Scan 3.** $B$ is empty and no buffer scan is performed.

The event sequence '$\diamond_P.?_Q.!_P.\diamond_Q$' in our initial buffer is delivered to $T_P$ unchanged. ◂

## B.2 Algorithm invariants

The invariants listed below ensure the correct handling of evt, dtc, rtd and messages by tracers. Lines 37, 51, and 60 in alg. 1, and lines 45 and 50 in alg. 3 include the main invariants below (respectively $I_{17}$, $I_{20}$, and $I_{19}$ in alg. 1 and $I_{22}$ in alg. 3). We elide the remaining invariants from algs. 1 and 3 in favour of presentation conciseness. As is the case with the invariants $I_{17}$, $I_{19}$, $I_{20}$, and $I_{22}$, our Erlang realisation of RIARC implements the elided ones as **assert** and **fail** statements. These invariants reason about general properties the tracer choreography should observe at all times. For instance, our invariants guarantee properties, such as, 'every trace event that is dispatched by the dispatch tracer eventually reaches the intended tracer', that 'the monitor choreography grows dynamically', and that 'redundant tracers are always garbage collected'. The invariants make use of three notions introduced in the main paper, which we recall for the benefit of readers.

▶ Note 10 (Tracers and messages).

- *Dispatch tracer*, sec. 3.2. A tracer that receives trace events meant to be handled by another tracer,
- *Forwarded message*, sec. 3.2. An evt or dtc message that is embedded in a rtd packet dispatched by a dispatch tracer,
- *Direct trace event*, sec. 3.3. An evt event that is not dispatched by a dispatched tracer but gathered from a SuS process via tracing. ◂

We organise invariants into two categories: the first describes properties of the tracer DAG topology, while the second focusses on tracer coordination and correct message delivery.

**Tracer choreography invariants**   Ensure that a DAG topology between tracers is always maintained by dynamic message routing.

$I_1$ A tracer *never* terminates unless its routing ($\Pi$) and traced-processes ($\Gamma$) maps are empty.

$I_2$ A tracer *never* adds a SuS PID that already exists in its traced-processes map $\Gamma$.

$I_3$ A tracer *never* removes an inexistent SuS PID from its traced-processes map $\Gamma$.

$I_4$ A tracer *always* acts on a $\diamond$ event by adding the spawned SuS PID to its traced-processes map $\Gamma$. *Requires invariant $I_2$ to hold.*

$I_5$ A tracer *always* acts on an $\star$ event by removing the SuS PID from its traced-processes map $\Gamma$. *Requires invariant $I_3$ to hold.*

$I_6$ A tracer *never* adds a next-hop that already exists in its routing map $\Pi$.

$I_7$ A tracer *never* removes an inexistent next-hop from its routing map $\Pi$.

$I_8$ A tracer *always* acts on a $\diamond$ event by adding a next-hop for the spawned SuS PID to its routing map $\Pi$. *Requires invariant $I_6$ to hold.*

$I_9$ A dispatch tracer that dispatches a $\diamond$ event *always* adds a next-hop for the spawned SuS PID to its routing map $\Pi$. *Requires invariant $I_6$ to hold.*

$I_{10}$ A tracer that forwards a $\diamond$ event *always* adds a next-hop for the spawned SuS PID to its routing map $\Pi$. *Requires invariant $I_6$ to hold.*

$I_{11}$ A dispatch tracer that dispatches a dtc acknowledgement *always* removes the corresponding next-hop for the detached SuS PID from its routing map $\Pi$. *Requires invariant $I_7$ to hold.*

$I_{12}$ A tracer that forwards a dtc acknowledgement *always* removes the corresponding next-hop for the detached SuS PID from its routing map $\Pi$. *Requires invariant $I_7$ to hold.*

**Message routing invariants**   Ensure that trace events are reported soundly to monitors.

$I_{13}$ A tracer *never* dispatches or forwards an evt or dtc message unless a route exists in its routing map $\Pi$. *Requires invariants $I_8$–$I_{10}$ to hold.*

$I_{14}$ A tracer in $\bullet$ mode *always* prioritises rtd packets until it switches to $\circ$ mode.

$I_{15}$ A tracer in $\bullet$ mode *always* transitions to $\circ$ mode only if all of the SuS PIDs in its traced-processes map $\Gamma$ are marked as $\circ$ or $\Gamma$ is empty.

$I_{16}$ The total amount of dtc requests a tracer issues is *always* equal to the sum of the number of SuS PIDs in its traced-processes map $\Gamma$ and the number of terminated SuS PIDs for the tracer. *Requires invariants $I_4$ and $I_5$ to hold.*

$I_{17}$ A tracer in $\circ$ mode *always* acts on a dtc request by dispatching it to the next-hop. *Requires invariants $I_{11}$ and $I_{13}$ to hold* (see line 37 in alg. 1).
If dispatching is not possible, the dtc request is incorrectly issued.

$I_{18}$ A tracer in $\circ$ mode *always* acts on a direct evt by analysing or dispatching it to the next-hop. *Requires invariant $I_{13}$ to hold.*

$I_{19}$ A tracer in $\circ$ mode *always* acts on a dispatched evt by forwarding it to the next-hop. *Requires invariant $I_{13}$ to hold* (see line 60 in alg. 1).
Analysing a dispatched evt in $\circ$ mode means that the tracer dequeued a priority event, violating invariant $I_{14}$.

$I_{20}$ A tracer in $\circ$ mode *always* acts on a dispatched dtc acknowledgement by forwarding it to the next-hop. *Requires invariants $I_{12}$ and $I_{13}$ to hold* (see line 51 in alg. 1).
Handling a dispatched dtc acknowledgement in $\circ$ mode means that the tracer dequeued a priority acknowledgement, violating invariant $I_{14}$.

$I_{21}$ A tracer in $\bullet$ mode *always* acts on a dispatched evt by analysing or forwarding it to the next-hop. *Requires invariant $I_{13}$ to hold.*

A tracer in • mode never dispatches events. Only tracers in ◦ mode can dispatch events, which are always direct events. Dispatching in • mode means that the tracer dequeued a non-priority event, violating invariant $I_{14}$.

**I$_{22}$** A tracer in • mode *always* acts on a dispatched dtc acknowledgement by handling or forwarding it to the next-hop. *Requires invariants $I_{12}$ and $I_{13}$ to hold* (see lines 45 and 50 in alg. 3).

A tracer in • mode never dispatches dtc acknowledgements. Only dispatch tracers in ◦ mode can dispatch dtc acknowledgements, which are always received from the tracers wishing to detach a SuS PID from the dispatch tracer. Dispatching in • mode means that the tracer dequeued a non-priority command, violating invariant $I_{14}$.

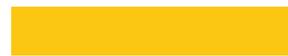

## C Appendix C: Empirical evaluation

App. C.1 details why existing benchmarking tools adopted in monolithic RV are inapplicable to our work. We use BenchCRV, which is tailored for setting up and building experiments that target RV for reactive systems; see apps. C.2 and C.3. The message numbering scheme BenchCRV employs in its master-worker models provides monitoring tools with a hook to implement assertions about trace events. We rely on this feature to ensure trace soundness in experiments. Our experiment set-up is summarised in app. C.4, along with a list of precautions in app. C.5. App. C.6 concludes with results supporting our arguments and conclusions in the main text.

### C.1 Benchmarking

Benchmarking is a standard method of gauging runtime overhead in software [103, 80, 36]. Established benchmarks such as SPECjvm2008 [136], DaCapo [28], Renaissance [122] ScalaBench [135]—developed for fine-tuning aspects of the JVM and actor libraries—are used by the RV community to assess the applicability of monitoring, *e.g.* see [116, 47, 46, 124, 30, 109, 81]. These frameworks rely on third-party off-the-shelf (OTS) programs to broaden and diversify benchmark coverage. *Synthetic benchmarks*, *e.g.* Savina [87], are an alternative way to perform benchmarking [34] and offer benefits over their OTS program-based analogues. For instance, parameters are used to induce variations in the core benchmark behaviour, enabling them to *reproduce* and control the *repeatability* of experiments. Interested readers are referred to [7] for a detailed account of the pros of synthetic benchmarking. All the benchmarking tools cited are *not* built with concurrency in mind, *e.g.* cannot generate high workloads that follow profiles typical in practice [7]. Along with synthetic benchmarking tools by the RV community [20, 68, 125, 22], the former ones gather metrics specific to *monolithic* batch-style programs (*e.g. execution slowdown*), which are orthogonal to reactive systems. These reasons make these tools inapplicable to our setting.

### C.2 BenchCRV workload parameters

BenchCRV generates workloads based on profiles observed in practice. A workload profile dictates how the master spreads its creation of worker processes along the loading timeline, specified by the parameter $t$ in seconds (s). The volume of workers per run is set via the parameter $n$. Every task the master allocates a worker consists of a *batch* of requests that the worker receives and echoes back to the master. The number of requests batched in one task is given by the parameter $w$. BenchCRV uses $w$ to generate different batch sizes for each worker to induce a modicum of variability in the master-worker models it generates. The actual batch size is generated within the range $w$ by drawing the number of work requests from a normal distribution with mean $\mu = w$ and standard deviation $\sigma = \mu \times 0.02$.

BenchCRV tool offers three load profiles.

**Steady** models scenarios where the SuS operates under stable conditions. The Steady workload is modelled on homogeneous Poisson distribution with *rate* $\lambda$, which specifies the mean number of workers created per second along the loading timeline with the duration $t = \lceil n/\lambda \rceil$.

**Pulse** models scenarios where the SuS experiences gradually rising and falling loads. The Pulse workload is configured by the *spread* parameter $\eta$, which determines how slowly or sharply the load increases as it nears its peak, halfway along $t$. Pulses are modelled on a Normal distribution with $\mu = t/2$ and $\sigma = \eta$.

| Param | Description |
|---|---|
| $n$ | Total number of worker processes per experiment |
| $w$ | Total number of requests per worker task |
| $t$ | Load timeline (inapplicable for Steady workload) |

**(a)** Master-worker model parameters

| Param | Description |
|---|---|
| $\lambda$ | Steady workload rate |
| $\eta$ | Pulse workload spread |
| $\pi$ | Burst workload pinch |
| $\Pr(send)$ | Probability master issues a work request |
| $\Pr(recv)$ | Probability master dequeues a work response |

**(b)** Workload and reactiveness parameters

**Table 4** BenchCRV configurable parameters for generating master-worker models and workloads

**Burst** models scenarios where the SuS is stressed due to load spikes. The Burst workload is configured by the *pinch* parameter $\pi$, which controls the concentration of the initial load burst. Bursts are modelled on a Log-normal distribution with $\mu = \ln(m^2/\sqrt{p^2+m^2})$ and $\sigma = \sqrt{\ln(1+p^2/m^2)}$.

Tbl. 4 summarises the parameters used to generate master-worker models (4a) and workloads (4b). Fig. 13 shows examples of the Steady, Pulse, and Burst workloads for a loading timeline of $t = 100$. These benchmarks are set with $n = 500\text{k}$ workers and $w = 100$ work requests per batch. The Steady workload is configured with $\lambda = 5\text{k}$, Pulse with $\eta = 25$, and Burst with $\pi = 100$.

Systems respond to load at different rates, *e.g.* due to the computational demand of tasks, IO, *etc.* BenchCRV simulates such phenomena via the parameters $\Pr(send)$ and $\Pr(recv)$. $\Pr(send)$ controls the probability that the master allocates requests to workers; $\Pr(recv)$ determines the probability that work responses received by the master are dequeued and acknowledged. Sending and receiving are *turn-based* and modelled on a Bernoulli trial [121]. The master picks a worker from its Work queue. It then draws a random number $X$ from a uniform distribution on the interval [0,1] and sends a work request when the Bernoulli trial succeeds, *i.e.*, $X \leq \Pr(send)$. The master decrements the work request counter for that worker and keeps sending requests to the same worker by drawing the next $X$ until the Bernoulli trial fails, *i.e.*, $X > \Pr(send)$, or the request counter reaches 0. If a Bernoulli trial fails on the first request-sending attempt, the worker misses its turn, and the next worker in the Work queue is picked. The master dequeues work responses it receives from workers using the scheme described. It repeatedly dequeues one response per successful Bernoulli trial, *i.e.*, $X \leq \Pr(recv)$, until the trial fails or the Receive queue is empty. The master signals workers to terminate once it acknowledges their work responses.

The developers of BenchCRV establish that adjusting $\Pr(send) = \Pr(recv) = 0.9$ yields SuS models that emulate *realistic* web-server response times. We use these recommended values in our experiments of sec. 5. Readers are referred to [7] for details.

### C.3 BenchCRV messaging model

The master-worker models that BenchCRV generates use a simple protocol to track the work requests allotted to different workers. Workers are initialised with IDs, which we denote by the placeholder *Id*, which enable the master to track the progress of *tasks* assigned. Each worker task comprises a sequence of work requests, *NumReqs*. The value of *NumReqs* for all workers is initially set to the value of the batch parameter $w$; see tbl. 4a. Work requests

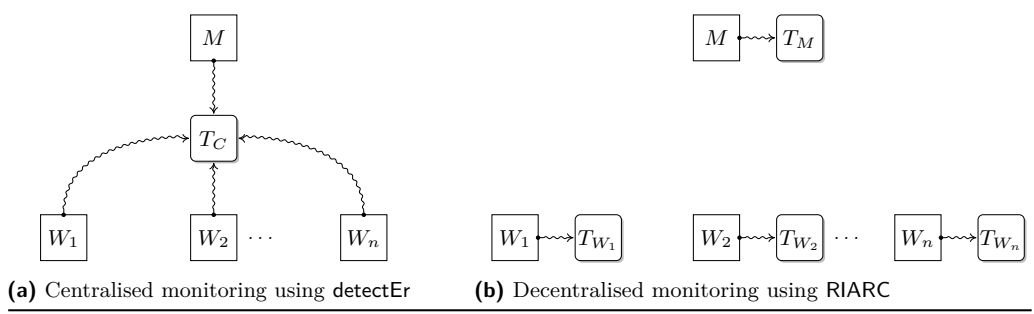

**(a)** Centralised monitoring using detectEr   **(b)** Decentralised monitoring using RIARC

**Figure 12** Centralised and RIARC monitoring arrangement on the master $M$ and workers $W_i$

in a task are assigned a unique sequence number, *ReqNum*, where $1 \leq \textit{ReqNum} \leq \textit{NumReqs}$, that identifies each request sent to a worker. The master process relies on *ReqNum* to determine when a task assigned to a particular worker is completed. A worker task completes when $\textit{ReqNum} = \textit{NumReqs}$, whereupon the master sends a special termination message to the worker. The triple $\langle \textit{Id}, \textit{ReqNum}, \textit{NumReqs} \rangle$ used in BenchCRV uniquely identifies work requests and responses in the system. BenchCRV relies on four messages to emulate work between the master and worker processes:

- $\langle \textit{Pid}_\text{M}, \langle \text{chunk}, \langle \textit{Id}, \textit{ReqNum}, \textit{NumReqs} \rangle \rangle \rangle$. Work request message that the master sends to the worker.
- $\langle \textit{Pid}_\text{M}, \langle \text{term}, \langle \textit{Id}, \textit{ReqNum}, \textit{NumReqs} \rangle \rangle \rangle$. Termination message that the master sends to the worker once the task is complete, *i.e.,* when $\textit{ReqNum} = \textit{NumReqs}$.
- $\langle \textit{Pid}_\text{W}, \langle \text{ack}, \langle \textit{Id}, \textit{ReqNum}, \textit{NumReqs} \rangle \rangle \rangle$. Work response message that the worker sends to the master.
- $\langle \textit{Pid}_\text{W}, \langle \text{end}, \langle \textit{Id}, \textit{ReqNum}, \textit{NumReqs} \rangle \rangle \rangle$. Completion message that the worker sends to the master when the last work request in a task is processed, *i.e.,* when $\textit{ReqNum} = \textit{NumReqs}$.

## C.4 Experiment set-up

Our empirical evaluation of sec. 5 configures benchmarks to monitor the master process and each worker that the master spawns. Fig. 12 overviews the arrangements of centralised and RIARC monitoring; inline monitoring follows that of fig. 1a. Inline monitoring uses the tool of [3, 4] to instrument the master and worker components in BenchCRV *statically*. The resulting modified code is then run in benchmarks. Centralised and RIARC monitoring rely on the EVM tracing to gather events without modifying the BenchCRV code. Our centralised monitoring benchmarks utilise detectEr [75, 16, 17, 15, 73, 40] to collectively instrument the master and every worker process with one central monitor. This central monitor, labelled $T_C$ in fig. 12a, analyses all the trace events gathered. The benchmarks set up with RIARC monitoring instrument the master and worker processes with identical monitor replicas, as illustrated in fig. 12b.

| Platform | Concurrency | Schedulers | Workers $n$ | Request batch $w$ | $\approx$ Messages | $\approx$ Messages/s |
|---|---|---|---|---|---|---|
| $P_E$ | $C_H$ | 4 | 100k | 100 | 20M | 162k |
| $P_G$ | $C_H$ | 16 | 500k | 100 | 100M | 218k |
| | $C_M$ | | 5k | 10k | 100M | 382k |

**Table 5** Benchmark configurations and message throughput at *maximum* Steady workloads

Tbl. 5 summarises all our experiment configurations from sec. 5.2. The table includes the mean throughput of work request and response messages exchanged between the master and worker processes under the Steady workload at its maximum. This maximum workload is at 100k workers for the high concurrency scenario $C_H$ on platform $P_E$, and at 500k workers for the high $C_H$ and at 5k workers for the moderate concurrency scenario $C_M$ on platform $P_G$. It is worth underscoring that the high and moderate concurrency settings used on platform $P_G$ yield an approximate number of messages in the master-worker models generated by BenchCRV. However, the throughput of 328k messages/s generated by $C_M$ is ≈76% higher than that of $C_H$ at 218k messages/s. This gap in throughput stems from the task batch size $w$, which controls the number of requests the master issues to each worker. $C_H$ and $C_M$ assess two facets of inline, centralised, and RIARC instrumentation:

**Stress handling** $C_H$ stresses each instrumentation method by inducing intense concurrency. The master provokes stress by spawning large numbers of workers ($n=500$k) continually during benchmark runs. Combined with the short worker lifespan due to modest request processing ($w=100$), this induces constant dynamic changes in the master-worker model. Intense concurrency tests the ability of RIARC to reorganise the tracer DAG topology and how this affects runtime overhead.

**Throughput handling** $C_M$ studies how instrumentation copes with high message throughput. The master creates comparatively fewer workers ($n=1$k), which engage in computationally long tasks ($w=100$k). Most workers are spawned in the first stages of benchmark runs and produce master-worker models exhibiting milder concurrency where workers terminate less frequently. Milder concurrency tests how RIARC operates in stabler conditions and how the infrequent trace event routing and tracer reconfigurations affect runtime overhead. Sec. 5.5 shows that inline and RIARC monitoring deliver similar results in these scenarios.

We reshape the stress and throughput factors described using the Steady, Pulse, and Burst workload profiles (see app. C.2). This variation increases our benchmark coverage and, in turn, the generality of our conclusions drawn from the results. Fig. 13 visualises the Steady, Pulse, and Burst workloads for the high concurrency scenario $C_H$ with 500k workers for each of the *ten* benchmark runs we use in experiments.

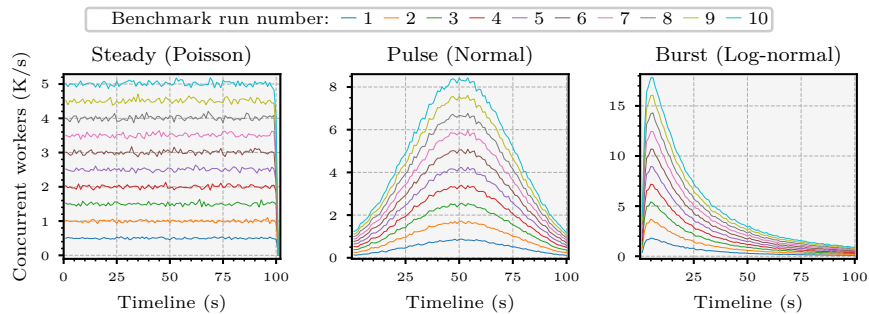

**Figure 13** Steady, Pulse and Burst workloads distributions of 500k workers sustained for 100s

## C.5  Precautions

The following precautions minimise the biases in our benchmarks and enhance the repeatability of our empirical evaluation presented in sec. 5.

### C.5.1  Repeatability

Data variability affects the repeatability of experiments [69]. The coefficient of variation (CV) [57], *i.e.,* the ratio of the standard deviation $\sigma$ to the mean $\bar{x}$, can be used to empirically establish the minimum number of experiment repetitions needed to obtain representative data. We denote this number by the variable $m$. The CV is calulated using $\text{CV} = \sigma/\bar{x}$.

We choose the minimum value of $m$ for our experiments as follows. First, we calculate the CV for the *first* batch of experiments for an initial number of repetitions $m$. This result, $cv$, is then compared to the CV calculation for the *next* batch of experiment repetitions, $m'$. The value $m'$ increments the number of benchmark repetitions to take by some batch offset value $b$, *i.e.,* $m' \leftarrow m + b$. We denote the CV obtained from the new calculation over $m'$ repetitions as $cv'$. The value $cv$ is subtracted from $cv'$: if the difference is sufficiently small for some error threshold $\epsilon$, the former number of repetitions, $m$, is selected. Otherwise, we repeat this procedure, setting $cv \leftarrow cv'$ and calculating the *new* CV value, $cv'$, for the next batch increment, $m'' \leftarrow m' + b$. Crucially, the condition $(cv' - cv) < \epsilon$ must hold for *all* the variables measured in the experiment before $m$ can be fixed. We perform these calculations to determine the number of benchmark repetitions used in sec. 5.

We also seed the Erlang pseudorandom number generator to minimise the data variability between experiments. Fixing the randomisation seed replicates the same workloads in all our experiments, making them repeatable. The upshot is that it requires fewer benchmark repetitions before the response time, memory consumption, and scheduler utilisation gathered by BenchCRV converge to an acceptable CV. Note that fixing the seed still permits our master-worker models to enjoy a degree of variability, which stems from the interleaved execution of processes due to scheduling.

### C.5.2  Centralised and decentralised monitoring

RIARC projects the global trace into partitions that reflect the *local* execution at SuS processes. It exploits the natural tree relationship induced by process spawning to create trace partitions, as sec. 2.1 remarks. By contrast, centralised monitoring gathers process events as one *global* trace sequence capturing the overall SuS behaviour. Existing work [47, 126] shows how a global trace can be efficiently sliced to recover trace partitions via a technique called parametric trace slicing (PTS). PTS generates the same local view of the SuS process execution induced by RIARC. Our centralised monitoring set up with detectEr employs PTS.

Its implementation consists of a specialised singleton monitor that *dynamically* demultiplexes the incoming stream of trace events. The projection relies on the PID carried by trace events, *i.e.,* $e.\imath_{\text{S}}$ in tbl. 1a of sec. 2.1, to direct them to corresponding local monitors. PTS enables us to reuse the monitors from our benchmarks with inline and RIARC monitoring. One crucial benefit of monitor reuse is that the *same* RV analysis logic is executed by the outline, inline, and RIARC monitors in our experiments, eliminating biases. The central monitor maintains a *monitor map* indexed by this PID to access the associated monitors efficiently and delegate the RV analysis. Our central monitor implementation ensures that every local monitor is created when needed and removed when its RV analysis completes. This measure guarantees the lowest possible overhead and does not bias our results against centralised monitoring.

The function AnalyseEvt($\varsigma_{\text{M}}$,$e$) conducts the RV analysis. AnalyseEvt takes a monitor signature, $\varsigma_{\text{M}}$, and reduces it by repeatedly applying it to the next event $e$ from a sequence of trace events. Each application, $\varsigma_{\text{M}}(e_i)$, returns the *new* monitor state $\varsigma'_{\text{M}}$, which is used for the next reduction, $\varsigma'_{\text{M}}(e_{i+1})$, and so forth. AnalyseEvt *stops* reducing $\varsigma_{\text{M}}$ when one of two conditions hold:

**Verdict flag** signals that the RV monitor *accepts* or *rejects* the behaviour of the SuS process based on the events analysed. We refer interested readers to [21, 15, 73] for an introduction to RV monitoring.

**End of partition** informs the RV monitor that there are *no* further trace events to analyse for the SuS process. The end of the partition is marked by the $\star$ event.

Either condition terminates the RV analysis, whereupon the monitor becomes stale. Sec. 3.6 overviews how stale monitors are disposed of when tracers are garbage collected.

In our empirical experiments, we use the sequence numbers carried by BenchCRV work request and response messages to ensure trace soundness; see app. C.3. Our specialised monitor signature $\varsigma_{\text{M}}$ maintains an internal offset to assert the trace event number, *ReqNum*, expected next. Monitors also confirm that the trace is reported in its entirety. We rely on *NumReqs*, which is used by BenchCRV worker processes to detect that all the work request messages from their respective batches are delivered to them. These basic checks guarantee that the trace event sequences monitors receive are *complete* and *consistent* per def. 1.

### C.6 Further results

We include further data plots supporting our conclusions of sec. 5.

### C.6.1 Monitoring overhead

Fig. 14 shows the overhead induced by centralised, inline, and RIARC monitoring. Charts include the overhead for the three monitoring methods under the Pulse workload to complete our findings from sec. 5.4.2. We recall that the *runtime monitoring* overhead combines the instrumentation and slowdown due to the RV analysis. Sec. 5.3 establishes this RV slowdown at ≈5µs per analysed trace event in our experiments. The slowdown stems from the runtime checking that our monitors perform to ensure that the trace event sequences reported by the instrumentation are sound, def. 1; see also app. C.5.2.

As fig. 8 from sec. 5.4.2, fig. 14 demonstrates that centralised monitoring crashes in our experiments (marked by ✗ in plots) when the Pulse workload is applied. The dumps recovered from crashes indicate that centralised monitoring fails for the reasons given in sec. 5.4.2. These plots also confirm that inline and RIARC monitoring are not afflicted by the ≈5µs RV analysis slowdown. We emphasise that RIARC induces almost comparable latency to inline monitoring even under the Pulse workload. Fig. 14 (top right, middle) puts the latency at 212ms for inline monitoring *vs.* 538ms for RIARC at a peak Pulse workload of 1.7k workers/s. The difference of 326ms between the two methods is lower than the 454ms gap calculated for the Burst workload in sec. 5.4.2.

The plots in fig. 14 (bottom) exhibit high scheduling utilisation: a byproduct of the limited number of scheduling threads (4) available on the edge-case platform $P_{\text{E}}$. Our plots in app. C.6.2 for experiments conducted on the general-case platform $P_{\text{G}}$ show that the scheduler utilisation is drastically reduced when using 16 scheduling threads. This reduction is exhibited even under the maximum workloads of ≈200M trace events, which is five times higher than the ≈40M workload used in fig. 14. Inline, and in particular, RIARC monitoring, benefit from the added scheduling capacity to scale accordingly. Centralised monitoring does not exhibit this behaviour; see app. C.6.2 for details.

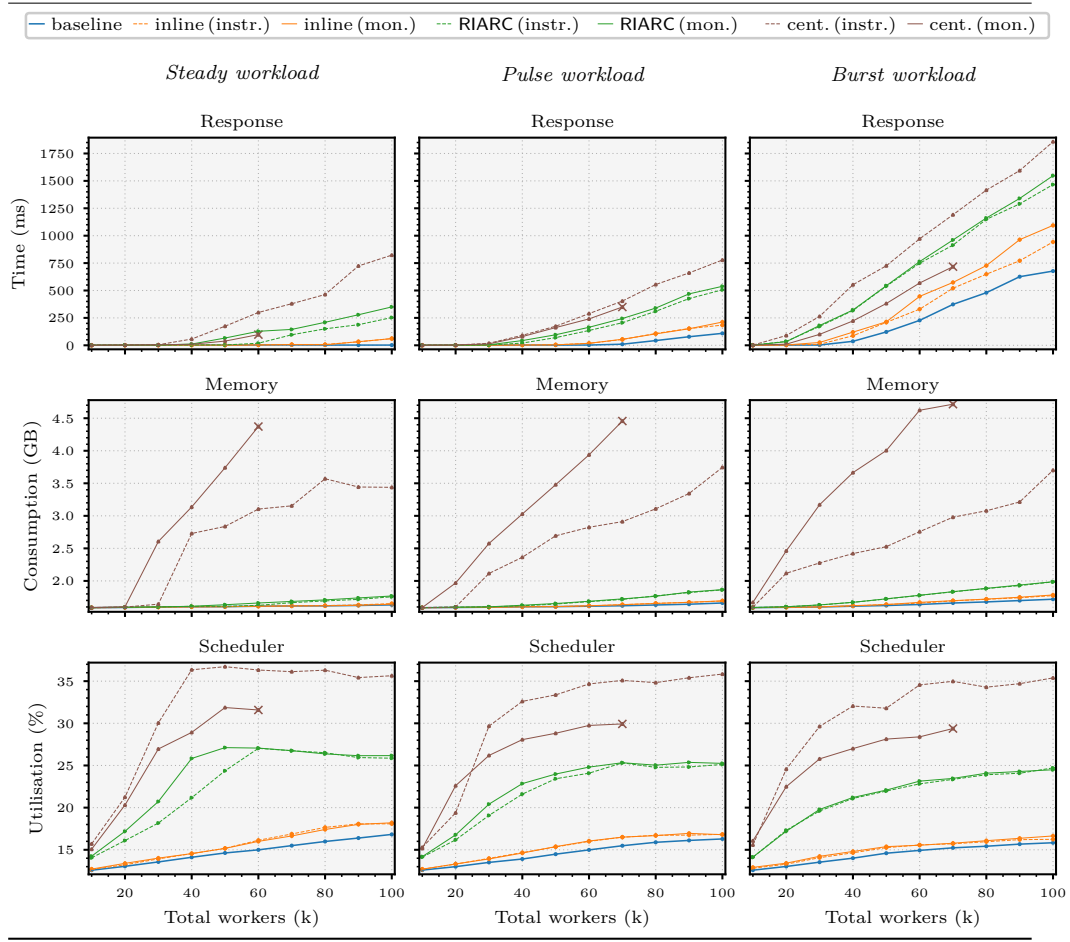

**Figure 14** Instrumentation and RV monitoring overhead gap (*high* workload, 100k workers)

## C.6.2 Scaled set-up

Our experiments on platform $P_E$ study how centralised, inline, and RIARC monitoring behave in edge-case situations where the memory is constrained, and the possibility of parallelism is limited; see app. C.6.1. The next set of experiments confirms that the same behaviour observed on platform $P_E$ for the three monitoring methods is preserved in general cases. These benchmarks are conducted on the general-case platform $P_G$ and use $n = 500$k workers, $w = 100$ requests per worker, and 16 scheduling threads.

Fig. 15 completes our view of instrumentation and runtime monitoring overhead given in fig. 8 from sec. 5.4.2. The memory consumption and scheduler utilisation plots of fig. 15 (bottom) magnify the bottleneck that afflicts centralised monitoring in fig. 8 of sec. 5.4.2. In the latter benchmarks taken on the edge-case platform $P_E$ with 100k workers, centralised monitoring plateaus to a mean scheduler utilisation of $\approx 31.8\%$ at the $\approx 50$k workers mark before eventually crashing. By comparison, the plots in fig. 15 show this to be at $\approx 4.7\%$ at the

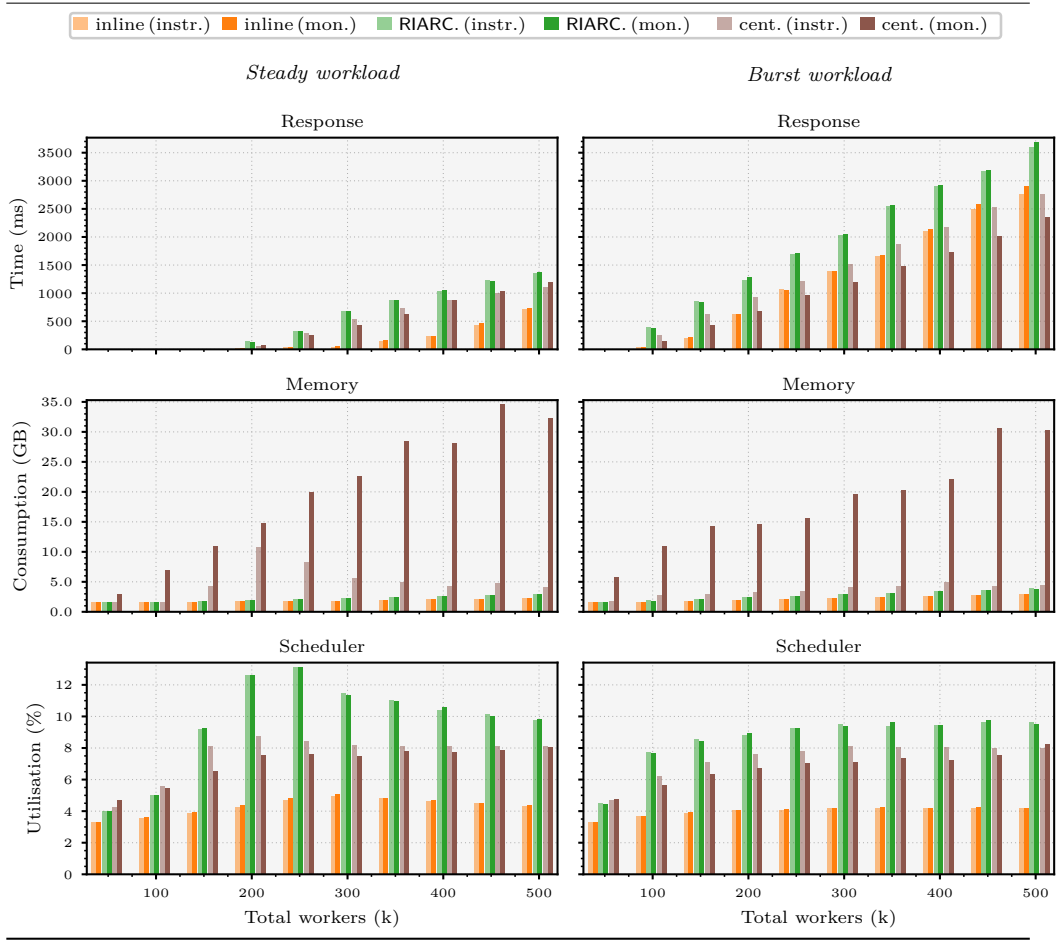

**Figure 15** Instrumentation and RV monitoring overhead gap (*high* workload, 500k workers)

*same* workload of 50k workers. This drop in scheduler utilisation for centralised monitoring stems from two reasons. First, the central monitor is limited in its use of the scheduling resources offered by platform $P_G$ due to the sequential processing of trace event messages. Second, the mean scheduler utilisation in this set-up is calculated over 16 scheduling threads.

Sec. 5.4.2 reports higher scheduler utilisation values on the edge-case platform $P_E$ because the EVM scheduling is limited to 4 threads; processes on $P_G$ are spread across more schedulers. The added parallelism gained through the extra 12 scheduling threads on platform $P_G$ permits workers to increase the message throughput in the corresponding master-worker models. For instance, the throughput of 162k messages/s with 100k workers under the Steady workload is raised to 218k messages/s in the benchmarks using 500k workers; refer to tbl. 5. This higher message throughput exacerbates the stress on the central monitor. We emphasise that the absence of crashes in the plots of fig. 15 is attributable to the considerable memory provided by the general-case platform $P_G$ rather than by the ability of centralised monitoring to cope with high workloads. Fig. 15 indicates that the continued increase in memory consumption eventually leads to failure when the memory capacity is exceeded.

Inline and RIARC monitoring enjoy the ample resources of platform $P_E$, scaling accordingly. This scalability manifests as conservative memory consumption and higher scheduler

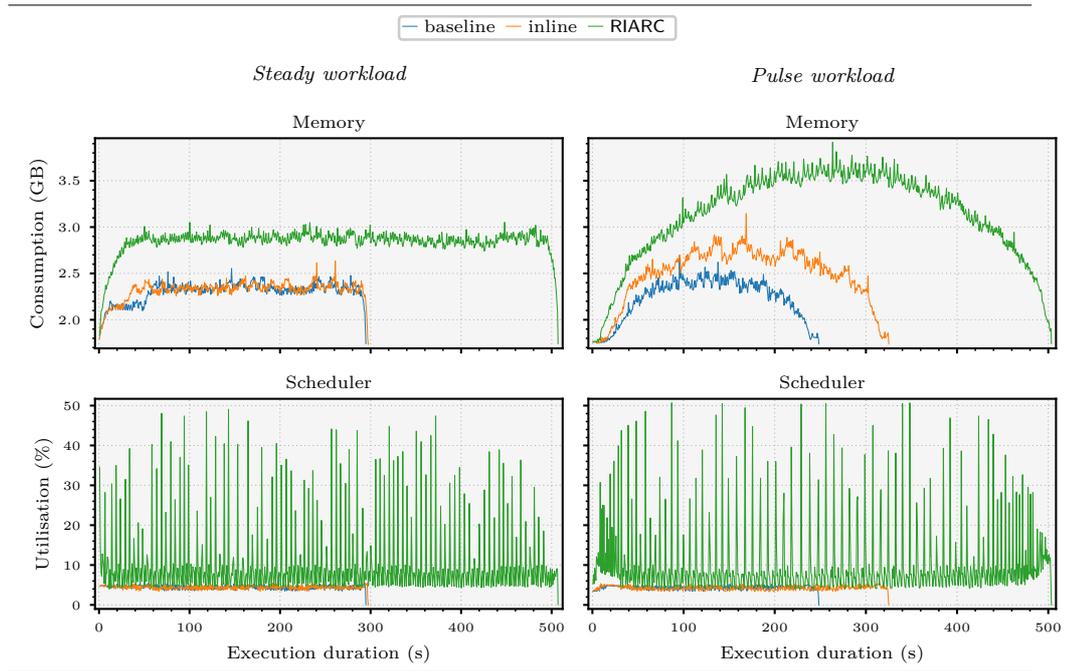

▮ **Figure 16** Inline and RIARC monitoring resource usage (*high* workload, 500k workers)

utilisation. Readers may notice the response time gains of centralised monitoring over inline and RIARC monitoring in fig. 15. We attribute this to very different reasons. The RV analysis slowdown causes the response time degradation in the case of inline monitoring. The latency overhead RIARC induces on our master-worker models is a byproduct of outline monitors, which compete for the same pool of scheduling threads used by worker processes. Under fair execution [137], workers reside in the EVM waiting queues for longer periods, impacting their ability to respond to work requests promptly. Fig. 8 in sec. 5.4.2 exhibits analogous behaviour. We conjecture that the response time for RIARC monitoring drastically improves in less extreme scenarios to those used for our benchmarks, which instrument *every* worker process in the model (see sec. 5.3).

### C.6.3 Resource usage

Sec. 5.4.3 gives an alternative view that studies the overall monitoring overhead—from the point of SuS launch until monitors complete their RV analysis. We supplement those results, showing that centralised monitoring is not scalable, whereas inline and RIARC monitoring leverage the extended processing capacity provided by the general-case platform $P_G$.

Fig. 16 complements fig. 9 in sec. 5.4.3, showing that inline and RIARC monitoring display elastic behaviour under Pulse workloads, too. Figs. 17 and 18 put the *same* plots of figs. 9 and 16 into the context of centralised monitoring. The former plots attest to the vast amounts of memory centralised monitoring consumes. They also highlight its lack of elasticity, where the memory consumption patterns are insensitive to the workload profile applied.

The sequential operation of the central monitor protracts the time taken for the RV analysis to complete. Such delays may render centralised monitoring inapplicable to cases where the RV set-up depends on timely detections, as in online monitoring. For instance, the benchmark runs captured in fig. 17 respectively take $\approx 862\%$ and $\approx 843\%$ longer to finish executing under the Steady and Burst workloads, when compared to the baseline system.

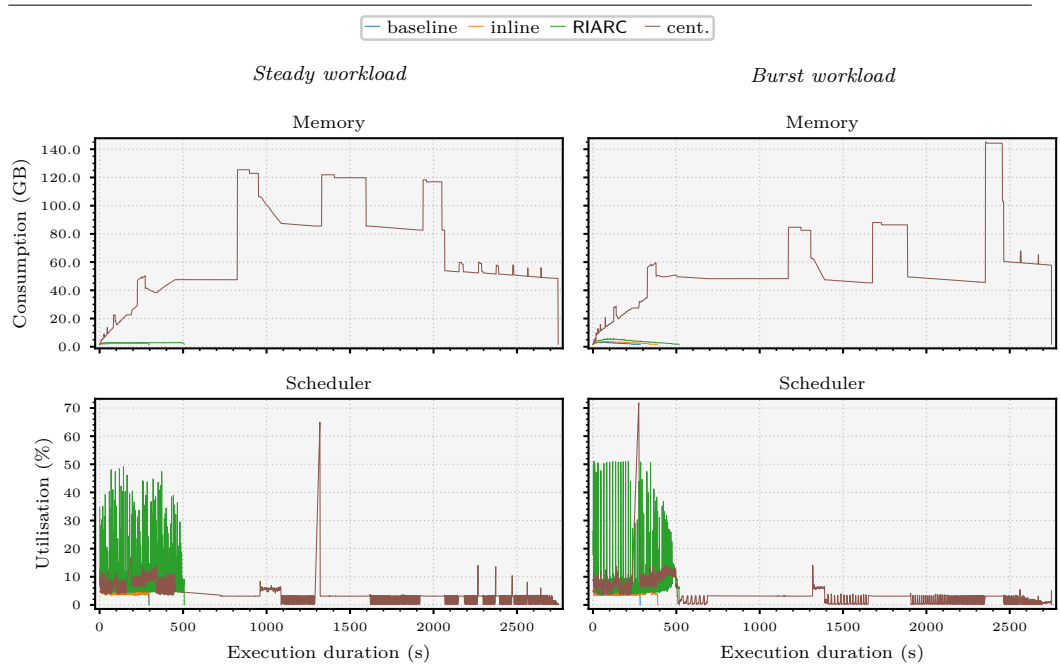

**Figure 17** Centralised, inline, and RIARC monitoring resource usage (*high* workload, 500k workers)

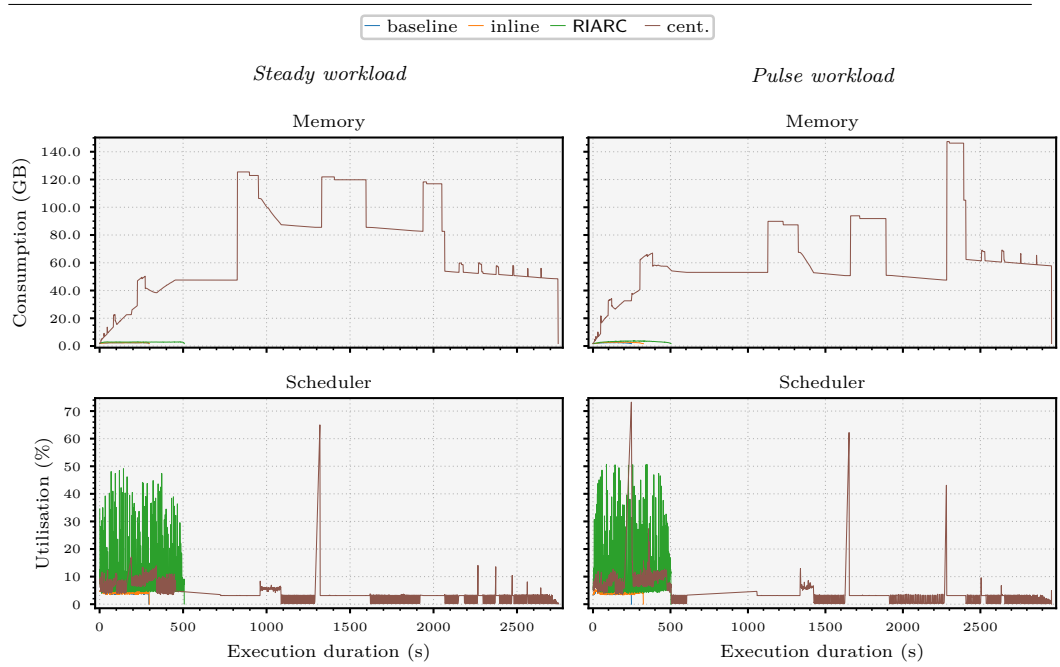

**Figure 18** Centralised, inline, and RIARC monitoring resource usage (*high* workload, 500k workers)

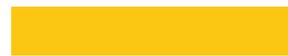

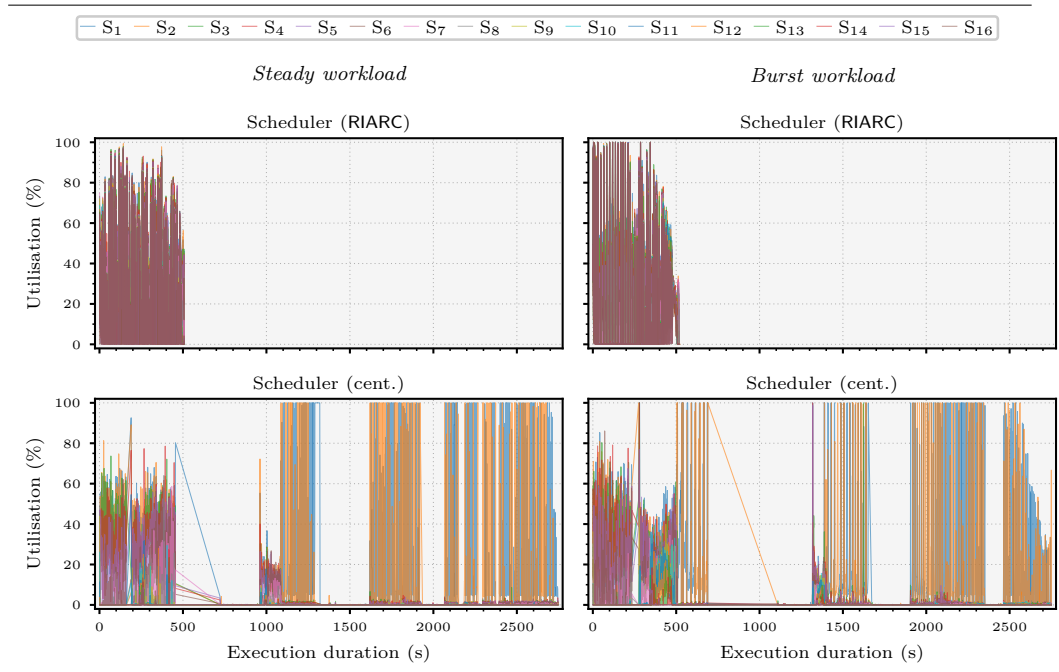

**Figure 19** Centralised, inline, and RIARC monitoring scheduler load (*high* workload, 500k workers)

Inline and RIARC monitoring terminate quicker under the same workloads. Inline monitoring registers an execution duration overhead of $\approx 1\%$ and $\approx 31\%$ w.r.t. baseline system in fig. 17 (bottom). RIARC monitoring prolongs the execution further, at $\approx 73\%$ and $\approx 85\%$ under the Steady and Pulse workloads. Fig. 18 for the Pulse workload shows analogous behaviour.

Fig. 9 of sec. 5.4.3 and fig. 16 unify the scheduler utilisation values by averaging over the 16 scheduler threads used in our general-case benchmarks on $P_G$. Scheduler oscillations with high peaks suggest simultaneous use of the scheduling threads. The absence of peaks in figs. 17 and 18 (bottom) for centralised monitoring results from the single-threaded monitor that cannot utilise other unoccupied EVM threads. Fig. 19 records the load on the individual EVM scheduling threads ($S_1$ to $S_{16}$) for the centralised and RIARC monitoring benchmark runs of fig. 17. The scheduler plots indicate *even load* distribution amongst the available threads for RIARC (top) under the Steady and Burst workloads. Even load distribution is consistent with the mean scheduler utilisation plots shown in fig. 17 for RIARC monitoring. By contrast, the load distribution for centralised monitoring in fig. 19 (bottom) becomes principally concentrated on scheduler threads $S_1$ and $S_2$ once the master and worker processes terminate. This behaviour is responsible for the *right skew* (*i.e.,* the right 'tail') in the scheduler utilisation plots of figs. 17 and 18 (bottom), which prolongs the execution of our centralised monitoring benchmarks.

### C.6.4 Moderate concurrency systems

Tbl. 3 in sec. 5.5 summarises the percentage overhead due to inline and RIARC monitoring w.r.t. the baseline system under the Steady and Burst workloads. These results are given on the general-case platform $P_G$ at *maximum* workloads with 500k workers (high concurrency, $C_H$) and 5k workers (moderate concurrency, $C_M$). Fig. 20 plots the results of *all* ten benchmark runs. The master process in our $C_H$ spawns substantially more worker processes than the master on $C_M$ in each corresponding benchmark run. These differences make the experiments on $C_H$ and $C_M$ incomparable in the number of processes created in a benchmark. For this reason, we use the benchmark run number (*x*-axis) to compare the overhead measured on $C_H$ and $C_M$ in fig. 20. We recall that the benchmarks on $C_H$ and $C_M$ generate an approximate volume of trace event messages.

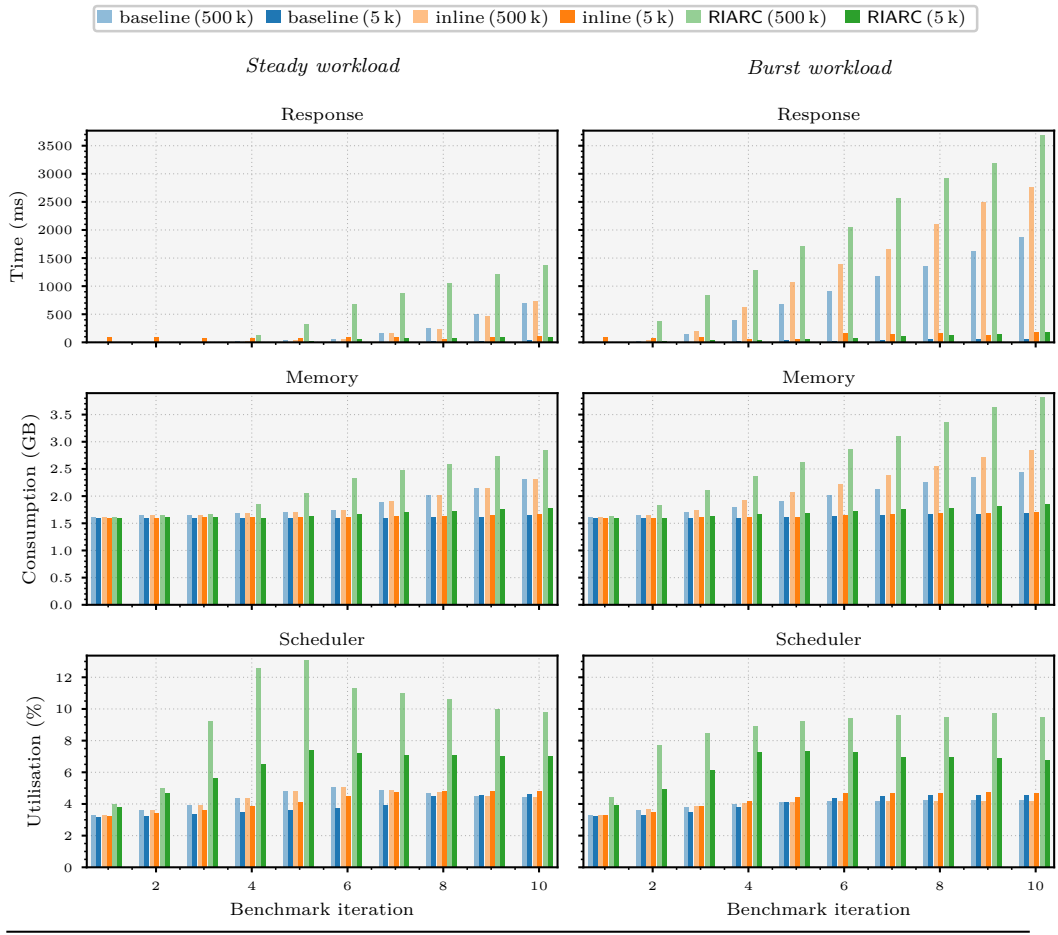

**Figure 20** Inline and RIARC monitoring overhead gap (*high/moderate* workload, 500k/5k workers)

Fig. 20 (bottom) registers negligible changes in scheduler utilisation between $C_M$ and $C_H$ for inline monitoring. Inline monitoring reduces its consumption of memory in our experiments with $C_M$. We attribute this to the lower number of workers BenchCRV creates relative to the models with $C_H$. This change lowers the strain on the master process induced by the constant spawning of workers throughout benchmark runs, which shrinks the memory footprint of the generated master-worker models. RIARC benefits from these moderately-sized master-worker models, as the memory consumption plots in fig. 20 (middle) indicate. However, most of the memory gains RIARC shows ensue from the fewer trace event routing and tracer reconfigurations it needs to perform compared to our experiments with concurrency scenario $C_H$. As a result, inline and RIARC monitoring consume comparable amounts of memory. RIARC recruits more scheduler capacity, $\approx 6.4\%$ *vs.* $\approx 4.2\%$ of inline monitoring under both the Steady and Burst workloads. This slight $\approx 2.2\%$ increase in scheduler utilisation enables RIARC to optimise the latency, bringing it *on par* with the latency induced by inline monitoring.